\documentclass[12pt]{article}
\begin{document}

\title{Gauging Newton's Law}
\author{James T. Wheeler\thanks{%
Department of Physics, Utah State University, Logan, Utah 84322} \thanks{%
Electronic mail: jwheeler@cc.usu.edu}}
\maketitle

\begin{abstract}
We derive both Lagrangian and Hamiltonian mechanics as gauge theories of
Newtonian mechanics. Systematic development of the distinct symmetries of
dynamics and measurement suggest that gauge theory may be motivated as a
reconciliation of dynamics with measurement. Applying this principle to
Newton's law with the simplest measurement theory leads to Lagrangian mechanics,
while use of conformal measurement theory leads to Hamilton's
equations. 
\end{abstract}

\section{Introduction}

Recent progress in field theory, when applied to classical physics, reveals
a previously unknown unity between various treatments of mechanics.
Historically, Newtonian mechanics, Lagrangian mechanics and Hamiltonian
mechanics evolved as distinct formulations of the content of Newton's second
law. Here we show that Lagrangian and Hamiltonian mechanics both arise as
local gauge theories of Newton's second law.

While this might be expected of Lagrangian mechanics, which is, after all,
just the locally coordinate invariant version of Newton's law, achieving
Hamiltonian mechanics as a gauge theory is somewhat surprising. The reason
it happens has to do with a new method of gauging scale invariance called 
\textit{biconformal gauging.} The study of biconformal gauging of Newtonian
mechanics serves a dual purpose. First, it sheds light on the meaning in
field theory of biconformal gauging, which has already been shown to have
symplectic structure and to lead to a satisfactory relativistic gravity
theory. Second, we are now able to see a conceptually satisfying unification
of Hamiltonian mechanics with its predecessors..

Beyond these reasons for the study, we find a hint of something deeper. The
biconformal gauging of Newton's law actually leads to something midway
between configuration space and phase space -- a $6$-dimensional symplectic
space in which motion is described by Hamilton's equations, but which does
not change size with increasing numbers of particles. While many of the
mathematical properties of Hamiltonian dynamics continue to hold, we are
offered a tantalizing glimpse of a new possibility -- perhaps this $6$-dimensional space is the proper ``arena'' for both classical and quantum
physics. In our conclusion, we revisit this conjecture.

Although the present article does not dwell on relativistic biconformal
spaces $per$ $se,$ we give a brief account of their history and properties.
The story starts with conformal gauge theories, which are notable for
certain pathologies: (1) the requirement for an invariant action in $2n$
dimensions to be of $n^{th}$ order in the curvature and/or the requirement
for auxiliary fields to write a linear action, and (2) the presence of
unphysical size changes. The existence of a second way to gauge the
conformal group was first explored by Ivanov and Niederle \cite
{Ivanov&Niederle}, who were led to an eight dimensional manifold by gauging
the conformal group of a four dimensional spacetime. They restricted the
dependence on the extra four dimensions to the minimum needed for
consistency with conformal symmetry. Later, Wheeler \cite{NewConfGauging},
generalizing to arbitrary dimensions, $n$, defined the class of biconformal
spaces as the result of the $2n$-dim gauging without imposing constraints,
showing it to have symplectic structure and admit torsion free spaces
consistent with general relativity and electromagnetism. Wehner and Wheeler 
\cite{WW} went on to write the most general class of actions linear in the
biconformal curvatures, eliminating problems (1) and (2) above, and showing
that the resulting field equations lead to the Einstein field equations.

For the purposes of the classical gauge theory considered here, it is
sufficient at the start to know that the full relativistic picture looks
even better: general relativity arises in a natural way from an action
principle linear in the biconformal curvatures. Unlike previous conformal
gauge theories, the biconformal gauging may be formulated in a uniform way
in any dimension, does not lead to unphysical size changes, and does not
require auxiliary fields. For further details, see \cite{NewConfGauging} and 
\cite{WW}. In the conclusion to this paper we comment briefly on the
phase-space-like interpretation of biconformal spaces.

In the next two sections, we make some observations regarding dynamical
laws, measurement theory and symmetry, then describe the global $ISO(3)$
symmetry of Newton's second law and the global $SO(4,1)$ symmetry of
Newtonian measurement theory. In Sec. 4, we give two ways to make these
different dynamical and measurement symmetries agree. After briefly
describing our method of gauging in Sec. 5, we turn to the actual gauging of
Newtonian mechanics. In Sec. 6 we show that the $ISO(3)$ gauge theory leads
to Lagrangian mechanics, while in Sec. 7 we show that biconformal gauging of
the $SO(4,1)$ symmetry leads to Hamiltonian dynamics, including at the end the case of multiple particles. In the penultimate section, we discuss an important question of interpretation, checking that there are no unphysical size changes.
Finally, we follow a brief summary with some observations about the possible
relevance to biconformal spaces and quantum physics.

\section{What constitutes a physical theory?}

We focus now on two essential features of any physical theory: dynamics and
measurement. Understanding the role played by each of these will lead us to
a deeper understanding of symmetry and ultimately to some novel conjectures
about the arena for physical theory.

To clarify the difference between dynamics and measurement, we first look at
quantum theory where the dynamics and measurement theories are quite
distinct from one another. Indeed, the dynamical law of quantum mechanics is
the Schr\"{o}dinger equation, 
\[
\hat{H}\psi =i\hbar \frac{\partial \psi }{\partial t} 
\]
This equation gives the time evolution of a state, $\psi ,$ but the state
has no direct physical meaning. Given a state, we still require a norm or an
inner product on states, 
\[
\left\langle \psi \mid \psi \right\rangle =\int_{V}\psi ^{*}\psi \ d^{3}x 
\]
to produce anything measurable. In addition, auxiliary rules for
interpretation are needed. Thus, the quantum norm above is interpreted as
the probability of finding the particle characterized by the state $\psi $
in the volume $V.$ Additional rules govern measurement of the full range of
dynamical variables.

Now we return to identify these elements of Newtonian mechanics. Newtonian
mechanics is so closely tied to our intuitions about how things move that we
don't usually separate dynamics and measurement as conceptually distinct.
Still, now that we know what we are looking for it is not difficult. The
dynamical law, of course, is Newton's second law: 
\[
F^{i}=m\frac{dv^{i}}{dt} 
\]
which describes the time evolution of a position vector of a particle. The
measurement theory goes back to the Pythagorean theorem -- it is based on
the line element or vector length in Euclidean space: 
\begin{eqnarray*}
ds^{2} &=&dx^{2}+dy^{2}+dz^{2} \\
&=&\eta _{ij}dx^{i}dx^{j} \\
\mathbf{v}\cdot \mathbf{w} &=&\eta _{ij}v^{i}w^{j}
\end{eqnarray*}
where 
\[
\eta _{ij}=\left( 
\begin{array}{ccc}
1 &  &  \\ 
& 1 &  \\ 
&  & 1
\end{array}
\right) 
\]
is the Euclidean metric in Cartesian coordinates. It is metric structure
that provides measurable numbers from the position vectors, forces and other
elements related by the dynamical equation. As we shall see below, there are
also further rules required to associate quantities computed from the
dynamical laws with numbers measured in the laboratory.

Once we have both a dynamical law and a measurement theory, we can begin
detailed exploration of the physical theory. Generally, this means analyzing
the nature of different interactions and making predictions about the
outcomes of experiments. For these two purposes -- studying interactions and
making predictions -- the most important tool is symmetry. The use of
symmetry for studying interactions follows from the techniques of gauge
theory, in which a dynamical law with a global symmetry is modified to be
consistent with a local symmetry of the same type. This procedure introduces
new fields into the theory, and these new fields generally describe
interactions. The use of symmetry for prediction relies on Noether's
theorem, which guarantees a conserved quantity corresponding to any
continuous symmetry. Once we have such a conserved quantity, we have an
immediate prediction: the conserved quantity will have the same value in the
future that it has now.

These three properties -- dynamics, measurement, and symmetry -- play a role
in every meaningful physical theory.

Once again, quantum theory provides a convenient example. Both the dynamical
law and the measurement theory make certain multiples of the wave function
equivalent. The dynamical law is linear, hence consistent with arbitrary
multiples of solutions. However, because of the derivatives involved in the
Schr\"{o}dinger equation, these multiples must be global, $\psi \rightarrow
A_{0}e^{i\varphi _{0}}\psi $. In contrast to this, we easily see that the
quantum norm is preserved by local multiples, as long as the multiple is a
pure phase: 
\[
\psi \rightarrow e^{\varphi \left( x\right) }\psi 
\]
Of course, $U\left( 1\right) $ gauge theory and the usual normalization of
the wave function provide one means of reconciling these differences in
symmetry. Notice that resolving the differences is accomplished by modifying
the symmetry of the dynamical equation to agree with that of the measurement
theory, both by restriction (fixing $A_{0}$ to normalize $\psi )$ and
extension (modifying the dynamical law to be consistent with local $U(1)$
transformations).

Gauging the $U(1)$ phase symmetry plays an extremely important role. By the
general procedure of gauging, we replace global symmetries by local ones,
and at the same time replace the dynamical law by one consistent with the
enlarged symmetry. Well-defined techniques are available for accomplishing
the required change in the dynamical laws. When the gauging procedure is
applied to the phase invariance of quantum field theory, the result is a
theory that includes electromagnetism. Thus, the gauging procedure provides
a way to systematically introduce interactions between particles -- forces.
In relativistic mechanics, gauging provides a successful theory of gravity
-- general relativity. As we shall see, gauging works well in the Newtonian
case too. Although we will not look for new interactions in the Newtonian gauge theory (these include gravity in the relativistic version), we will see that gauging leads directly to Hamilton's equations.

The symmetry of Newtonian mechanics is often taken to be the set of
transformations relating inertial frames. We can arrive at this conclusion
by asking what transformations leave the dynamical equation invariant. The
answer is that Newton's second law is invariant under any transformation of
the form 
\begin{eqnarray*}
x^{i} &\rightarrow &J_{\quad j}^{i}x^{j}+v^{i}t+x_{0}^{i} \\
F^{i} &\rightarrow &J_{\quad j}^{i}F^{j}
\end{eqnarray*}
where $J_{\quad j}^{i}$ is a constant, nondegenerate matrix and $v^{i}$ and $
x_{0}^{i}$ are constant vectors. A shift in the time coordinate and time
reversal are also allowed. However, not all of these are consistent with the
measurement theory. If we ask which of the transformations above also
preserve the Pythagorean norm, we must further restrict $J_{\quad j}^{i}$ to
be orthogonal. Newtonian mechanics is thus invariant under 
\begin{eqnarray*}
x^{i} &\rightarrow &O_{\quad j}^{i}x^{j}+v^{i}t+x_{0}^{i} \\
F^{i} &\rightarrow &O_{\quad j}^{i}F^{j} \\
t &\rightarrow &t+t_{0}
\end{eqnarray*}
While this brief argument leads us to the set of orthogonal inertial frames,
it is not systematic. Rather, as we shall see, this is a conservative
estimate of the symmetries that are possible.

In the next sections, we treat the symmetries of Newtonian mechanics in a
more systematic way. In preparation for this, recall that in the quantum
phase example, we both restricted and extended the dynamical law to
accommodate a symmetry of the measurement theory, but arriving at the
inertial frames for the Newtonian example we only restricted the symmetry of
the dynamical law. This raises a general question. When the dynamical law
and measurement theory have different symmetries, what do we take as the
symmetry of the theory? Clearly, we should demand that the dynamical
equations and the measurement theory share a common set of symmetry
transformations. If there is a mismatch, we have three choices:

\begin{enumerate}
\item  Restrict the symmetry to those shared by both the dynamical laws and
the measurement theory.

\item  Generalize the measurement theory to one with the same symmetry as
the dynamical law.

\item  Generalize the dynamical equation to one with the same symmetry as
the measurement theory.
\end{enumerate}

We annunciate and apply the \textit{Goldilocks Principle: }Since we
recognize that symmetry sometimes plays an important predictive role in
specifying possible interactions, option \#1 is \textit{too small}. It is
unduly restrictive, and we may miss important physical content. By contrast,
the symmetry of measurement is \textit{too large} for option \#2 to work --
inner products generally admit a larger number of symmetries than dynamical
equations. Option \#3 is \textit{just right}: there are general techniques
for enlarging the symmetry of a dynamical equation to match that of a
measurement theory. Indeed, this is precisely what happens in gauge
theories. The extraordinary success of gauge theories may be because they
extend the dynamical laws to agree with the maximal information permitted
within a given measurement theory.

We will take the point of view that the largest possible symmetry is
desirable, and will therefore always try to write the dynamical law in a way
that respects the symmetry of our measurement theory. This leads to a novel
gauging of Newton's law. In the next section we look in detail at two
symmetries of the second law: the usual Euclidean symmetry, $ISO(3),$ and
the $SO(4,1)$ conformal symmetry of a modified version of Newton's law. Each
of these symmetries leads to an interesting gauge theory.

\section{Two symmetries of classical mechanics}

In this section we first find the symmetry of Newton's second law, then find
the symmetry of Newtonian measurement theory.

\subsection{Symmetry of the dynamical equation}

Newton's second law 
\begin{equation}
\mathbf{F}=m\frac{d\mathbf{v}}{dt}  \label{Newton's Second Law}
\end{equation}
has several well-known symmetries. The point symmetries leaving eq.(\ref
{Newton's Second Law}) invariant are derived in Appendix 1. The result is
that two allowed coordinate systems must be related by a constant,
inhomogeneous, general linear transformation, together with a shift (and
possible time reversal) of $t:$%
\begin{eqnarray}
\tilde{x}^{m} &=&J_{\;\;n}^{m}x^{n}+v_{0}^{m}t+x_{0}^{m}
\label{Inertial frames} \\
\tilde{t} &=&t+t_{0}  \label{Time translation} \\
\tilde{F}^{m} &=&J_{\;\;n}^{m}F^{n}  \label{Force transform}
\end{eqnarray}
where $J_{\;\;n}^{m}$ is any constant, non-degenerate matrix, $v_{0}^{m}$
and $x_{0}^{m}$ are arbitrary constant vectors, and $t_{0}$ is any real
constant.

Notice that, setting $e^{\lambda }=\left| \det \left( J_{\;\;n}^{m}\right)
\right| $, Newton's second law transforms covariantly with respect to
rescaling of units. Both sides of the dynamical law transform by an overall
factor of $e^{-2\lambda }.$

Eqs.(\ref{Inertial frames}-\ref{Force transform}) gives a $16$-parameter
family of transformations: nine for the independent components of the $%
3\times 3$ matrix $J,$ three for the boosts $v_{0}^{i},$ three more for the
arbitrary translation, $x_{0}^{m},$ and a single time translation. The
collection of all of these coordinate sets constitutes the maximal set of
inertial systems. This gives us the symmetry of the dynamical law.

\subsection{Symmetry of Newtonian measurement theory}

Newtonian measurement theory begins with the Pythagorean theorem as embodied
in the line element and corresponding vector product 
\begin{eqnarray}
ds^{2} &=&dx^{2}+dy^{2}+dz^{2}=\eta _{ij}dx^{i}dx^{j} \\
\mathbf{v}\cdot \mathbf{w} &=&\eta _{ij}v^{i}w^{j}
\end{eqnarray}
The line element is integrated to find lengths of curves, while the dot
product lets us find components of vectors by projecting on a set of basis
vectors. But to have a line element or an inner product is not enough to
have a theory of measurement. We must be specific about how the numbers
found from the inner product relate to numbers measured in the laboratory.

Suppose we wish to characterize the magnitude of a displacement vector, $%
\mathbf{x,}$ separating two particles by using the Euclidean inner product, 
\begin{equation}
\Vert \mathbf{x}\Vert ^{2}=\eta _{ij}x^{i}x^{j}  \label{Length of a vector}
\end{equation}
The result, to be meaningful, must still be expressed in some set of units,
say, meters or centimeters. The fact that either meters or centimeters will
do may be expressed by saying that we work with an equivalence class of
metrics differing by a positive multiplier. Thus, if we write eq.(\ref
{Length of a vector}) for the length of $\mathbf{x}$ in meters, then to give
the length in centimeters we must write 
\[
\Vert \mathbf{x}\Vert ^{2}=10^{4}\eta _{ij}x^{i}x^{j} 
\]
We regard these two metrics as equivalent, and indeed, all metrics of the
form 
\[
g_{ij}=e^{2\lambda \left( x\right) }\eta _{ij} 
\]
The factor $e^{2\lambda \left( x\right) }$ is called a conformal factor; two
metrics which differ by a conformal factor are conformally equivalent.

The symmetry group which preserves conformal equivalence classes of metrics
is the global conformal group, locally isomorphic to $O(4,1).$ The conformal
group is comprised of the following transformations: 
\[
y^{i}=\left\{ 
\begin{array}{ll}
O_{\;\;j}^{i}x^{j} & Orthogonal\ transformation \\ 
x^{i}+a^{i} & Translation \\ 
e^{\lambda }x^{i} & Dilatation \\ 
\frac{x^{i}+x^{2}b^{i}}{1+2b\cdot x+b^{2}x^{2}} & Special\ conformal\
transformation
\end{array}
\right. \hspace{0.25in} 
\]
The first three of these are familiar symmetries. We now discuss each of the
conformal symmetries, and the relationship between the $SO(4,1)$ symmetry of
classical measurement theory and the $ISO(3)$ symmetry of the dynamical law.

\subsection{Relationship between the dynamical and measurement symmetries}

As expected, there are some simple relationships between the symmetries of
Newton's second law and the symmetries of the Euclidean line element.
Indeed, if we restrict to global conformal transformations, the first three
-- orthogonal transformations, translations, and dilatations -- all are
allowed transformations to new inertial frames. We only need to restrict the
global general linear transformations $J_{\;\;n}^{m}$ of eqs.(\ref{Inertial
frames}) and (\ref{Force transform}) to orthogonal, $O_{\;\;n}^{m}$ for
these to agree, while the $v_{0}^{m}t+x_{0}^{m}$ part of eq.(\ref{Inertial
frames}) is a parameterized global translation.

For global dilatations we see the invariance of Newton's second law simply
because the units on both sides of the equation match: 
\begin{eqnarray*}
\left[ \mathbf{F}\right] &=&\frac{kg\cdot m}{s^{2}} \\
\left[ m\mathbf{a}\right] &=&kg\cdot \frac{m}{s^{2}}
\end{eqnarray*}
The dilatation corresponds to $e^{-2\lambda }=\left| \det \left(
J_{\;\;n}^{m}\right) \right| $ in eqs.(\ref{Inertial frames}) and (\ref
{Force transform}). Notice that the conformal transformation of units
considered here is completely different from the conformal transformations
(or renormalization group transformations) often used in quantum field
theory. The present transformations are applied to \textit{all} dimensionful
fields, and it is impossible to imagine this simple symmetry broken. By
contrast, in quantum field theory only certain parameters are renormalized
and there is no necessity for dilatation invariance.

To make the effect of dilatations more transparent, we use fundamental
constants to express their units as $\left( length\right) ^{k}$ for some
real number $k$ called the \textit{conformal\ weight}$.$ This allows us to
quickly compute the correct conformal factor. For example, force has weight $%
k=-2$ since we may write 
\[
\left[ \frac{1}{%TCIMACRO{\UNICODE[m]{0x127}}
%BeginExpansion
\rlap{\protect\rule[1.1ex]{.325em}{.1ex}}h%
%EndExpansion
c}\mathbf{F}\right] =\frac{1}{l^{2}} 
\]
The norm of this vector then transforms as 
\[
\left\| \frac{1}{%TCIMACRO{\UNICODE[m]{0x127}}
%BeginExpansion
\rlap{\protect\rule[1.1ex]{.325em}{.1ex}}h%
%EndExpansion
c}\mathbf{F}\right\| \rightarrow e^{-2\lambda }\left\| \frac{1}{%
%TCIMACRO{\UNICODE[m]{0x127}}
%BeginExpansion
\rlap{\protect\rule[1.1ex]{.325em}{.1ex}}h%
%EndExpansion
c}\mathbf{F}\right\| 
\]
With this understanding, we see that Newton's law, eq.(\ref{Newton's Second
Law}), transforms covariantly under global dilatations. With 
\begin{eqnarray*}
\left[ \frac{1}{%TCIMACRO{\UNICODE[m]{0x127}}
%BeginExpansion
\rlap{\protect\rule[1.1ex]{.325em}{.1ex}}h%
%EndExpansion
c}\mathbf{F}\right] &=&\frac{1}{l^{2}} \\
\left[ \frac{m\mathbf{v}}{%TCIMACRO{\UNICODE[m]{0x127}}
%BeginExpansion
\rlap{\protect\rule[1.1ex]{.325em}{.1ex}}h%
%EndExpansion
}\right] &=&\frac{1}{l} \\
\left[ \frac{1}{c}\frac{d}{dt}\right] &=&\frac{1}{l}
\end{eqnarray*}
the second law, 
\begin{equation}
\left( \frac{1}{%TCIMACRO{\UNICODE[m]{0x127}}
%BeginExpansion
\rlap{\protect\rule[1.1ex]{.325em}{.1ex}}h%
%EndExpansion
c}\right) \mathbf{F}=\frac{1}{c}\frac{d}{dt}\left( \frac{m\mathbf{v}}{%
%TCIMACRO{\UNICODE[m]{0x127}}
%BeginExpansion
\rlap{\protect\rule[1.1ex]{.325em}{.1ex}}h%
%EndExpansion
}\right)
\end{equation}
has units $\left( length\right) ^{-2}$ throughout: 
\begin{equation}
\left[ \left( \frac{1}{%TCIMACRO{\UNICODE[m]{0x127}}
%BeginExpansion
\rlap{\protect\rule[1.1ex]{.325em}{.1ex}}h%
%EndExpansion
c}\right) \mathbf{F}\right] =\left[ \left( \frac{1}{c}\frac{d}{dt}\right)
\left( \frac{m\mathbf{v}}{%TCIMACRO{\UNICODE[m]{0x127}}
%BeginExpansion
\rlap{\protect\rule[1.1ex]{.325em}{.1ex}}h%
%EndExpansion
}\right) \right] =\frac{1}{l^{2}}
\end{equation}
Under a global dilatation, we therefore have 
\begin{equation}
e^{-2\lambda }\left( \frac{1}{%TCIMACRO{\UNICODE[m]{0x127}}
%BeginExpansion
\rlap{\protect\rule[1.1ex]{.325em}{.1ex}}h%
%EndExpansion
c}\right) \mathbf{F}=e^{-\lambda }\frac{1}{c}\frac{d}{dt}\left( e^{-\lambda }%
\frac{m\mathbf{v}}{%TCIMACRO{\UNICODE[m]{0x127}}
%BeginExpansion
\rlap{\protect\rule[1.1ex]{.325em}{.1ex}}h%
%EndExpansion
}\right) =e^{-2\lambda }\frac{1}{c}\frac{d}{dt}\left( \frac{m\mathbf{v}}{%
%TCIMACRO{\UNICODE[m]{0x127}}
%BeginExpansion
\rlap{\protect\rule[1.1ex]{.325em}{.1ex}}h%
%EndExpansion
}\right)
\end{equation}
Newton's law is therefore globally dilatation covariant, of conformal weight 
$-2.$

The story is very different for special conformal transformations. These
surprising looking transformations are translations in inverse coordinates.
Defining the inverse to any coordinate $x^{i}$ as 
\[
y^{i}=-\frac{x^{i}}{x^{2}} 
\]
we find the general form of a special conformal transformation by inverting,
translating by an arbitrary, constant vector, $-b^{i},$ then inverting once
more: 
\[
x^{i}\rightarrow -\frac{x^{i}}{x^{2}}\rightarrow -\frac{x^{i}}{x^{2}}%
-b^{i}\rightarrow q^{i}=\frac{x^{i}+x^{2}b^{i}}{1+2b^{i}x_{i}+b^{2}x^{2}} 
\]
The inverse is given by the same sequence of steps, with the opposite sign
of $b^{i},$ applied to $q^{i},$ i.e., 
\[
x^{i}=\frac{q^{i}-q^{2}b^{i}}{1-2q^{i}b_{i}+q^{2}b^{2}} 
\]
As we show in Appendix 2, the transformation has the required effect of
transforming the metric according to 
\begin{equation}
\eta _{ab}\rightarrow \left( 1-2b^{i}x_{i}+b^{2}x^{2}\right) ^{-2}\eta _{ab}
\end{equation}
and is therefore conformal. This time, however, the conformal factor is not
the same at every point. These transformations are nonetheless global
because the parameters $b^{i}$ are constant -- letting $b^{i}$ be an
arbitrary function of position would enormously enlarge the symmetry in a
way that no longer returns a multiple of the metric.

In its usual form, Newton's second law is \textit{not} invariant under
global special conformal transformations. The derivatives involved in the
acceleration do not commute with the position dependent transformation:

\begin{equation}
e^{-2\lambda \left( b,x\right) }\left( \frac{1}{%
%TCIMACRO{\UNICODE[m]{0x127}}
%BeginExpansion
\rlap{\protect\rule[1.1ex]{.325em}{.1ex}}h%
%EndExpansion
c}\right) \mathbf{F}\neq e^{-\lambda \left( x\right) }\frac{1}{c}\frac{d}{dt}
\left( e^{-\lambda \left( b,x\right) }\frac{\partial x^{i}}{\partial q^{j}}%
\frac{mv^{j}}{%TCIMACRO{\UNICODE[m]{0x127}}
%BeginExpansion
\rlap{\protect\rule[1.1ex]{.325em}{.1ex}}h%
%EndExpansion
}\right)  \label{Noncovariance of spec conf}
\end{equation}
and the dynamical law is not invariant.

\section{A consistent global symmetry for Newtonian mechanics}

Before we can gauge ``the'' symmetry of Newtonian mechanics, we face the
dilemma described in the second section: our measurement theory and our
dynamical equation have different symmetries. The usual procedure for
Newtonian mechanics is to restrict to the intersection of the two
symmetries, retaining only global translations and global orthogonal
transformations, giving the inhomogeneous orthogonal group, $ISO(3).$ This
group can then be gauged to allow local $SO(3)$ transformations. However, in
keeping with our (Goldilocks) principal of maximal symmetry, and noting that
the conformal symmetry of the measurement theory is larger than the
Euclidean symmetry of the second law, we will rewrite the second law with
global conformal symmetry, $O(4,1).$ The global conformal symmetry may then
be gauged to allow local $SO(3)\times R^{+}$ (homothetic) transformations.
In subsequent sections we will carry out both of these gaugings.

Our goal in this section is to write a form of Newton's second law which is
covariant with respect to global conformal transformations. To begin, we
have the set of global transformations 
\begin{eqnarray*}
y^{i} &=&O_{\quad j}^{i}x^{j} \\
y^{i} &=&x^{i}+a^{i} \\
y^{i} &=&e^{\lambda }x^{i} \\
y^{i} &=&\frac{x^{i}+x^{2}b^{i}}{1+2b\cdot x+b^{2}x^{2}}=\beta ^{-1}\left(
x^{i}+x^{2}b^{i}\right)
\end{eqnarray*}
As seen above, it is the derivatives that obstruct the full conformal
symmetry (see eq.(\ref{Noncovariance of spec conf})). The first three
transformations already commute with ordinary partial differentiation of
tensors because they depend only on the constant parameters $O_{\quad
j}^{i},a^{i}$ and $\lambda .$ After a special conformal transformation,
however, the velocity becomes a complicated function of position, and when
we compute the acceleration, 
\[
a^{i}=\frac{dv^{i}}{dt}=\frac{\partial y^{i}}{\partial x^{j}}\frac{d^2 x^{i}}{d t^2}+v^{k}\frac{\partial }{\partial x^{k}}\left( \frac{%
\partial y^{i}}{\partial x^{j}}v^{j}\right)
\]
the result is not only a terrible mess -- it is a different terrible mess
than what we get from the force (see Appendix 3). The problem is solved if
we can find a new derivative operator that commutes with special conformal
transformations.

The mass also poses an interesting problem. If we write the second law as 
\begin{equation}
\mathbf{F}=\frac{d}{dt}\left( m\mathbf{v}\right)
\end{equation}
we see that even ``constant'' scalars such as mass pick up position
dependence and contribute unwanted terms when differentiated 
\begin{eqnarray*}
m &\rightarrow &e^{-\lambda \left( x\right) }m \\
\partial _{i}m &\rightarrow &e^{-\lambda \left( x\right) }\partial
_{i}m-e^{-\lambda \left( x\right) }m\partial _{i}\lambda
\end{eqnarray*}
We can correct this problem as well, with an appropriate covariant
derivative.

To find the appropriate derivation, we consider scalars first, then vectors,
with differentiation of higher rank tensors following by the Leibnitz rule. For
nonzero conformal weight scalars we require 
\[
D_{k}s_{\left( n\right) }=\partial _{k}s_{\left( n\right) }+ns_{\left(
n\right) }\Sigma _{k} 
\]
while for vectors we require a covariant derivative of the form, 
\[
D_{k}v_{\left( n\right) }^{i}=\partial _{k}v_{\left( n\right)
}^{i}+v_{\left( n\right) }^{j}\Lambda _{jk}^{i}+nv_{\left( n\right)
}^{i}\Sigma _{k} 
\]

Continuing first with the scalar case, we easily find the required
transformation law for $\Sigma _{k}.$ Transforming $s_{\left( n\right) }$ we
demand covariance, 
\[
D_{k}^{\prime }s_{\left( n\right) }^{\prime }=\left( D_{k}s_{\left( n\right)
}\right) ^{\prime }
\]
where 
\[
D_{k}^{\prime }s_{\left( n\right) }^{\prime }=e^{-\lambda }\partial %
_{k}\left( e^{n\lambda }s_{\left( n\right) }\right) +n\left( e^{n\lambda
}s_{\left( n\right) }\right) \Sigma _{k}^{\prime }
\]
and 
\[
\left( D_{k}s_{\left( n\right) }\right) ^{\prime }=e^{n^{\prime }\lambda
}\left( D_{k}s_{\left( n\right) }\right) 
\]
Since derivatives have conformal weight $-1,$ we expect that\footnote{%
In field theory, the coordinates and therefore the covariant derivative are
usually taken to have zero weight, while dynamical fields and the metric
carry the dimensional information. In Newtonian physics, however, the
coordinate is also a dynamical variable, and must be assigned a weight.} 
\[
n^{\prime }=n-1
\]
Imposing the covariance condition, 
\begin{eqnarray*}
e^{-\lambda }\partial _{k}\left( e^{n\lambda }s_{\left( n\right) }\right)
+n\left( e^{n\lambda }s_{\left( n\right) }\right) \Sigma _{k}^{\prime }
&=&e^{n^{\prime }\lambda }\left( D_{k}s_{\left( n\right) }\right)  \\
e^{-\lambda }\left( s_{\left( n\right) }n\partial _{k}\lambda +\partial %
_{k}s_{\left( n\right) }\right) +ns_{\left( n\right) }\Sigma _{k}^{\prime }
&=&e^{-\lambda }\left( \partial _{k}s_{\left( n\right) }+ns_{\left( n\right)
}\Sigma _{k}\right)  \\
s_{\left( n\right) }n\partial _{k}\lambda +ne^{\lambda }s_{\left( n\right) }%
\Sigma _{k}^{\prime } &=&ns_{\left( n\right) }\Sigma _{k}
\end{eqnarray*}
or 
\[
\Sigma _{k}^{\prime }=e^{-\lambda }\left( \Sigma _{k}-\partial _{k}\lambda
\right) 
\]
Since we assume the usual form of Newton's law holds in some set of
coordinates, $\Sigma _{k}$ will be zero for these coordinate systems.
Therefore, we can take $\Sigma _{k}$ to be zero until we perform a special
conformal transformation, when it becomes $-ne^{-\lambda }\partial %
_{k}\lambda .$ Notice that since $\lambda $ is constant for a dilatation, $%
\Sigma _{k}$ remains zero if we simply change from furlongs to feet.

Since a special conformal transformation changes the metric from the flat
metric $\eta _{ij}$ to the conformal metric 
\begin{equation}
g_{ij}=e^{2\lambda \left( x\right) }\eta _{ij}=\beta ^{-2}\eta _{ij}
\label{Conformal metric}
\end{equation}
where 
\begin{equation}
\beta =1+2\mathbf{b\cdot x}+b^{2}x^{2}  \label{Conformal factor}
\end{equation}
we need a connection consistent with a very limited set of coordinate
transformations. This just leads to a highly restricted form of the usual
metric compatible Christoffel connection. From eq.(\ref{Conformal metric})
we compute immediately, 
\begin{eqnarray}
\Lambda _{jk}^{i} &=&\frac{1}{2}g^{im}\left(
g_{mj,k}+g_{mk,j}-g_{jk,m}\right)  \nonumber \\
&=&\eta ^{im}\left( \eta _{mj}\lambda _{,k}+\eta _{mk}\lambda _{,j}-\eta
_{jk}\lambda _{,m}\right)  \label{Connection}
\end{eqnarray}
where 
\[
\lambda _{,k}=-\beta ^{-1}\beta _{,k} 
\]
Notice that $\Lambda _{jk}^{i}$ has conformal weight $-1,$ and vanishes
whenever $b_{i}=0.$

We can relate $\Sigma _{k}$ directly to the special conformal connection $\Lambda _{jk}^{i}$. The trace of $\Lambda _{jk}^{i}$ is 
\[
\Lambda _{k}\equiv \Lambda _{ik}^{i}=3\lambda _{,k} 
\]
so that 
\begin{eqnarray*}
\Sigma _{k} &=&-\lambda ,_{k} \\
&=&-\frac{1}{3}\Lambda _{k}
\end{eqnarray*}
The full covariant derivative of, for example, a vector of conformal weight $%
n,$ may therefore be written as 
\begin{equation}
D_{k}v_{\left( n\right) }^{i}=\partial _{k}v_{\left( n\right)
}^{i}+v_{\left( n\right) }^{j}\Lambda _{jk}^{i}-\frac{n}{3}v_{\left(
n\right) }^{i}\Lambda _{k}  \label{Covariant derivative}
\end{equation}
where $\Lambda _{jk}^{i}$ is given by eq.(\ref{Connection}).

\subsection{Constant mass and conformal dynamics}

Extending the symmetry of classical mechanics to include special conformal
transformations introduces an unusual feature: even constants such as mass
may appear to be position dependent. But we are now in a position to say
what it means for a scalar to be constant. Since mass has conformal weight $%
-1,$ we demand 
\[
D_{k}m=\partial _{k}m+\frac{1}{3}\Lambda _{k}m=0 
\]
That is, constant mass now means covariantly constant mass.

This equation is always integrable because $\Lambda _{k}$ is curl-free, 
\[
\Lambda _{k,m}-\Lambda _{m,k}=3\left( \lambda _{,km}-\lambda _{,mk}\right)
=0 
\]
Integrating, 
\[
m=m_{0}e^{-\lambda } 
\]
Any set of masses, $\left\{ m_{\left( 1\right) },m_{\left( 2\right) },\ldots
,m_{\left( N\right) }\right\} ,$ in which each element satisfies the same
condition,

\[
D_{k}m_{\left( i\right) }=0,\qquad i=1,\ldots ,N
\]
gives rise to an invariant spectrum of measurable mass ratios, 
\[
M_{R}=\left\{ \frac{m_{1}}{m_{0}},\frac{m_{2}}{m_{0}},\ldots ,\frac{m_{N}}{%
m_{0}}\right\} 
\]
since the conformal factor cancels out. Here we have chosen $\frac{%
%TCIMACRO{\UNICODE[m]{0x127}}
%BeginExpansion
\rlap{\protect\rule[1.1ex]{.325em}{.1ex}}h%
%EndExpansion
}{m_{0}c}$ as our unit of length.

We can also write Newton's second law in an invariant way. The force is a
weight $-2$ vector. With the velocity transforming as a weight zero vector and the mass as a wehght $-1$ scalar, the time derivative of the momentum now requires a covariant derivative, 
\[
\frac{D\left(mv^{i}\right)}{Dt}=\frac{d }{dt} \left(mv^{i}\right)+ mv^{j} v^{k}\Lambda _{jk}^{i}+\frac{1}{3}mv^{i} v^{k}\Lambda _{k}
\]
Then Newton's law is 
\[
F^{i}=\frac{D}{Dt}\left( mv^{i}\right)
\]
To see how this extended dynamical law transforms, we check conformal
weights. The velocity has the dimensionless form 
\[
\frac{1}{c}\frac{dx^{i}}{dt} 
\]
The covariant derivative reduces this by one, so the acceleration has
conformal weight $-1.$ The mass also has weight $-1,$ while the force, as
noted above, has weight $-2.$ Then we have: 
\[
\tilde{F}^{i}=\frac{D}{D\tilde{t}}\left(\tilde{ m}\tilde{v}^{i}\right)
\]
The first term in the covariant time derivative becomes
\begin{eqnarray*}
\frac{d}{d\tilde{t}}\left(\tilde{ m}\tilde{v}^{i}\right) &=& e^{-\lambda}\frac{d}{dt}\left(e^{-\lambda}  m\frac{\partial y^{i}}{\partial x^{j}}v^{j}\right) \\
&=& e^{-2\lambda }\frac{\partial y^{i}}{\partial x^{j}}\frac{d}{dt}\left( mv^{j}\right)+e^{-\lambda}mv^{j}\frac{dx^{k}}{dt}\frac{\partial}{\partial x^{k}}\left(e^{-\lambda} \frac{\partial y^{i}}{\partial x^{j}}\right)
\end{eqnarray*}
The final term on the right exactly cancels the inhomogeneous contributions from 
$\Lambda _{jk}^{i}$ and $\Lambda _{k}$, leaving the same conformal factor and  Jacobian that multiply the force:
\[
e^{-2\lambda }\frac{\partial y^{i}}{\partial x^{j}}F^{j} = e^{-2\lambda }\frac{\partial y^{i}}{\partial x^{j}}\left(\frac{d\left(mv^{j}\right)}{d t}+
mv^{m} v^{k}\Lambda _{mk}^{j}+\frac{1}{3}mv^{j} v^{k}\Lambda _{k}\right) 
\]
The conformal factor and Jacobian cancel, so if the globally conformally covariant Newton's equation holds in one conformal frame, it holds in all conformal frames.

The transformation to the conformally flat metric 
\[
g_{ij}=e^{2\lambda }\eta _{ij}=\beta ^{-2}\eta _{ij} 
\]
does not leave the curvature tensor invariant. This only makes sense -- just
as we have an equivalence class of metrics, we require an equivalence class
of curved spacetimes. The curvature for $g_{ij}$ is computed in Appendix 4.

We now want to consider what happens when we gauge the symmetries associated
with classical mechanics. In the next section, we outline some basics of
gauge theory. Then in succeeding sections we consider two gauge theories
associated with Newtonian mechanics. First, we gauge the Euclidean $ISO(3)$
invariance of $F^{i}=ma^{i}$, then the full $O(4,1)$ conformal symmetry of $%
F^{i}=\frac{D}{Dt}\left( mv^{i}\right) $.

\section{Gauge theory}

Here we briefly outline the quotient group method of gauging a symmetry
group. Suppose we have a Lie group, $\mathcal{G}$, with corresponding Lie
algebra 
\[
\left[ G_{A},G_{B}\right] =c_{AB}^{\quad C}G_{C} 
\]
Suppose further that $\mathcal{G}$ has a subgroup $\mathcal{H}$, such that $%
\mathcal{H}$ itself has no subgroup normal in $\mathcal{G}$. Then the
quotient group $\mathcal{G}/\mathcal{H}$ is a manifold with the symmetry $%
\mathcal{H}$ acting independently at each point (technically, a fiber
bundle). $\mathcal{H}$ is now called the isotropy subgroup. The manifold
inherits a connection from the original group, so we know how to take $%
\mathcal{H}$-covariant derivatives. We may then generalize both the manifold
and the connection, to arrive at a class of manifolds with curvature, still
having local $\mathcal{H}$ symmetry. We consider here only the practical
application of the method. Full mathematical details may be found, for
example, in \cite{Kobayashi&Nomizu}.

The generalization of the connection proceeds as follows. Rewriting the Lie
algebra in the dual basis of 1-forms defined by 
\[
\left\langle G_{A},\mathbf{\omega }^{B}\right\rangle =\delta _{A}^{B} 
\]
we find the Maurer-Cartan equation for $\mathcal{G}$, 
\[
\mathbf{d\omega }^{C}=-\frac{1}{2}c_{AB}^{\quad C}\mathbf{\omega }^{A}\wedge 
\mathbf{\omega }^{B} 
\]
This is fully equivalent to the Lie algebra above. We consider the quotient
by $\mathcal{H}$. The result has the same appearance, except that now all of
the connection 1-forms $\mathbf{\omega }^{A}$ are regarded as linear
combinations of a smaller set spanning the quotient. Thus, if the Lie
algebra of $\mathcal{H}$ has commutators 
\[
\left[ H_{a},H_{b}\right] =c_{ab}^{\quad c}H_{c} 
\]
then the Lie algebra for $\mathcal{G}$ may be written as 
\begin{eqnarray*}
\left[ G_{\alpha },G_{\beta }\right] &=&c_{\alpha \beta }^{\quad \rho
}G_{\rho }+c_{\alpha \beta }^{\quad a}H_{a} \\
\left[ G_{\alpha },H_{a}\right] &=&c_{\alpha a}^{\quad \rho }G_{\rho
}+c_{\alpha a}^{\quad b}H_{b} \\
\left[ H_{a},H_{b}\right] &=&c_{ab}^{\quad c}H_{c}
\end{eqnarray*}
where $\alpha $ and $a$ together span the full range of the indices $A.$
Because $\mathcal{H}$ contains no normal subgroup of $\mathcal{G}$, the
constants $c_{\alpha a}^{\quad \rho }$ are nonvanishing for some $\alpha $
for all $a.$ The Maurer-Cartan structure equations take the corresponding
form 
\begin{eqnarray}
\mathbf{d\omega }^{\rho } &=&-\frac{1}{2}c_{\alpha \beta }^{\quad \rho }%
\mathbf{\omega }^{\alpha }\wedge \mathbf{\omega }^{\beta }-\frac{1}{2}%
c_{\alpha a}^{\quad \rho }\mathbf{\omega }^{\alpha }\wedge \mathbf{\omega }%
^{a}  \label{Structure eq 1} \\
\mathbf{d\omega }^{a} &=&-\frac{1}{2}c_{\alpha \beta }^{\quad a}\mathbf{%
\omega }^{\alpha }\wedge \mathbf{\omega }^{\beta }-\frac{1}{2}c_{\alpha
b}^{\quad a}\mathbf{\omega }^{\alpha }\wedge \mathbf{\omega }^{b}-\frac{1}{2}%
c_{bc}^{\quad a}\mathbf{\omega }^{b}\wedge \mathbf{\omega }^{c}
\label{Structure equations}
\end{eqnarray}
and we regard the forms $\mathbf{\omega }^{a}$ as linearly dependent on the $%
\mathbf{\omega }^{\alpha },$%
\[
\mathbf{\omega }^{a}=\omega _{\alpha }^{a}\mathbf{\omega }^{\alpha } 
\]
The forms $\mathbf{\omega }^{\alpha }$ span the base manifold and the $%
\mathbf{\omega }^{a}$ give an $\mathcal{H}$-symmetric connection.

Of particular interest for our formulation is the fact that eq.(\ref
{Structure eq 1}) gives rise to a covariant derivative. Because $\mathcal{H}$
is a subgroup, $\mathbf{d\omega }^{\rho }$ contains no term quadratic in $%
\mathbf{\omega }^{a},$ and may therefore be used to write 
\[
0=\mathbf{D\omega }^{\rho }\equiv \mathbf{d\omega }^{\rho }+\mathbf{\omega }%
^{\alpha }\wedge \mathbf{\omega }_{\alpha }^{\quad \rho } 
\]
with 
\[
\mathbf{\omega }_{\alpha }^{\quad \rho }\equiv \frac{1}{2}c_{\alpha \beta
}^{\quad \rho }\mathbf{\omega }^{\beta }-\frac{1}{2}c_{\alpha a}^{\quad \rho
}\mathbf{\omega }^{a} 
\]
This expresses the covariant constancy of the basis. As we shall see in our $%
SO(3)$ gauging, this derivative of the orthonormal frames $\mathbf{\omega }%
^{\rho }$ is not only covariant with respect to local $\mathcal{H}$
transformations, but also leads directly to a covariant derivative with
respect to general coordinate transformations when expressed in a coordinate
basis. This is the reason that general relativity may be expressed as both a
local Lorentz gauge theory and a generally coordinate invariant theory, and
it is the reason that Lagrangian mechanics with its ``generalized
coordinates'' may also be written as a local $SO(3)$ gauge theory.

Continuing with the general method, we introduce curvature two forms. These
are required to be quadratic in the basis forms $\mathbf{\omega }^{\alpha }$
only, 
\begin{eqnarray*}
\mathbf{R}^{a} &=&\frac{1}{2}R_{\quad \alpha \beta }^{a}\mathbf{\omega }%
^{\alpha }\wedge \mathbf{\omega }^{\beta } \\
\mathbf{R}^{\rho } &=&\frac{1}{2}R_{\quad \alpha \beta }^{\rho }\mathbf{%
\omega }^{\alpha }\wedge \mathbf{\omega }^{\beta }
\end{eqnarray*}
Then the modified connection is found by solving 
\begin{eqnarray*}
\mathbf{d\omega }^{\rho } &=&-\frac{1}{2}c_{\alpha \beta }^{\quad \rho }%
\mathbf{\omega }^{\alpha }\wedge \mathbf{\omega }^{\beta }-\frac{1}{2}%
c_{\alpha a}^{\quad \rho }\mathbf{\omega }^{\alpha }\wedge \mathbf{\omega }%
^{a}+\mathbf{R}^{\rho } \\
\mathbf{d\omega }^{a} &=&-\frac{1}{2}c_{\alpha \beta }^{\quad a}\mathbf{%
\omega }^{\alpha }\wedge \mathbf{\omega }^{\beta }-\frac{1}{2}c_{\alpha
b}^{\quad a}\mathbf{\omega }^{\alpha }\wedge \mathbf{\omega }^{b}-\frac{1}{2}%
c_{bc}^{\quad a}\mathbf{\omega }^{b}\wedge \mathbf{\omega }^{c}+\mathbf{R}%
^{a}
\end{eqnarray*}
and we may use any manifold consistent with this local structure.

This technique is used, for example, as a gauge approach to general
relativity by constructing a class of manifolds with local Lorentz
structure. One takes the quotient of the Poincar\'{e} group by the Lorentz
group. Our constructions of Newtonian theory will illustrate the method,
although we will not generalize to curved spaces or different manifolds. As
a result, the structure equations in the form of eqs.(\ref{Structure eq 1}, \ref{Structure equations}) describe the geometry and symmetry of our gauged dynamical law.

\section{A Euclidean gauge theory of Newtonian mechanics}

We begin by gauging the usual restricted form of the second law, using the
Euclidean group as the initial global symmetry. The familiar form (see
Appendix 5) of the Lie algebra of the Euclidean group, $iso(3),$ is 
\begin{eqnarray*}
\left[ J_{i},J_{j}\right] &=&\varepsilon _{ij}^{\quad k}J_{k} \\
\left[ J_{i},P_{j}\right] &=&\varepsilon _{ij}^{\quad k}P_{k} \\
\left[ P_{i},P_{j}\right] &=&0
\end{eqnarray*}
Using the quotient group method, we choose $so(3)$ as the isotropy subgroup.
Then introducing the Lie algebra valued 1-forms $\mathbf{\omega }^{i}$ dual
to $J_{i}$ and $\mathbf{e}^{i}$ dual to $P_{i}$ we write the Maurer-Cartan
structure equations 
\begin{eqnarray*}
\mathbf{d\omega }^{m} &=&-\frac{1}{2}c_{ij}^{\quad m}\mathbf{\omega }^{i}%
\mathbf{\omega }^{j}=-\frac{1}{2}\varepsilon _{ij}^{\quad m}\mathbf{\omega }%
^{i}\mathbf{\omega }^{j} \\
\mathbf{de}^{m} &=&-c_{ij}^{\quad m}\mathbf{\omega }^{i}\mathbf{e}%
^{j}=-\varepsilon _{ij}^{\quad m}\mathbf{\omega }^{i}\mathbf{e}^{j}
\end{eqnarray*}
Defining 
\[
\mathbf{\omega }^{k}=\frac{1}{2}\varepsilon _{\quad mn}^{k}\mathbf{\omega }%
^{mn} 
\]
these take a form familiar from general relativity, 
\begin{eqnarray}
\mathbf{d\omega }^{mn} &=&\mathbf{\omega }^{mk}\mathbf{\omega }_{k}^{\quad n}
\label{Spin connection} \\
\mathbf{de}^{m} &=&\mathbf{e}^{k}\mathbf{\omega }_{k}^{\quad m}
\label{Solder form}
\end{eqnarray}
with $\mathbf{\omega }^{mn}$ the spin connection and $\mathbf{e}^{m}$ the
dreibein. These equations are equivalent to the commutation relations of the
Lie algebra, with the Jacobi identity following as the integrability
condition $\mathbf{d}^{2}=0,$ i.e., 
\begin{eqnarray*}
\mathbf{d}^{2}\mathbf{\omega }^{mn} &=&\mathbf{d}\left( \mathbf{\omega }^{mj}%
\mathbf{\omega }_{j}^{\quad n}\right) =\mathbf{d\omega }^{mj}\mathbf{\omega }%
_{j}^{\quad n}-\mathbf{\omega }^{mj}\mathbf{d\omega }_{j}^{\quad n} \\
&\equiv &0 \\
\mathbf{d}^{2}\mathbf{e}^{m} &=&\mathbf{d}\left( \mathbf{\omega }_{\quad
k}^{m}\mathbf{e}^{k}\right) =\mathbf{d\omega }_{\quad k}^{m}\mathbf{e}^{k}-%
\mathbf{\omega }_{\quad k}^{m}\mathbf{de}^{k} \\
&\equiv &0
\end{eqnarray*}
Eqs.(\ref{Spin connection}) and (\ref{Solder form}) define a connection on a
three dimensional (flat) manifold spanned by the three 1-forms $\mathbf{e}%
^{m}.$ We take $\mathbf{\omega }^{mn}$ to be a linear combination of the $%
\mathbf{e}^{m}.$ This completes the basic construction.

The equations admit an immediate solution because the spin connection, $%
\mathbf{\omega }^{mn}$ is in involution. The $6$-dimensional group manifold
therefore admits coordinates $y^{i}$ such that 
\[
\mathbf{\omega }^{mn}=w_{\quad \quad \alpha }^{mn}\mathbf{d}y^{\alpha } 
\]
and there are submanifolds given by $y^{m}=const.$ On these 3-dimensional
submanifolds, $\mathbf{\omega }^{mn}=0$ and therefore 
\[
\mathbf{de}^{m}=\mathbf{\omega }_{\quad k}^{m}\mathbf{e}^{k}=0 
\]
with solution 
\[
\mathbf{e}^{m}=\delta _{\alpha }^{m}\mathbf{d}x^{\alpha } 
\]
for an additional three coordinate functions $x^{\alpha }.$ This solution
gives Cartesian coordinates on the $y^{\alpha }=const.$ submanifolds.
Identifying these manifolds as a copies of our Euclidean 3-space, we are now
free to perform an arbitrary rotation at each point.

Performing such local rotations on orthonormal frames leads us to general
coordinate systems. When we do this, the spin connection $\mathbf{\omega }%
^{mn}$ takes the more general, pure gauge, form 
\[
\mathbf{\omega }^{mn}=-\left( \mathbf{d}O_{\quad j}^{m}\right) \bar{O}^{jn} 
\]
where $O_{\quad j}^{m}\left( x\right) $ is a local orthogonal transformation
and $\bar{O}^{jn}\left( x\right) $ its inverse. Then $\mathbf{e}^{i}$
provides a general orthonormal frame field, 
\[
\mathbf{e}^{i}=e_{\alpha }^{\quad i}\mathbf{d}x^{\alpha } 
\]
The coefficients $e_{\alpha }^{\quad i}\left( x\right) $ may be determined
once we know $O_{\quad j}^{m}\left( x\right) .$

The second Maurer-Cartan equation gives us a covariant derivative as
follows. Expand any 1-form in the orthonormal basis, 
\[
\mathbf{v}=v_{i}\mathbf{e}^{i} 
\]
Then we define the covariant exterior derivative via 
\begin{eqnarray*}
\left( \mathbf{D}v_{i}\right) \mathbf{e}^{i} &=&\mathbf{dv} \\
&=&\mathbf{d}\left( v_{i}\mathbf{e}^{i}\right) \\
&=&\mathbf{d}v_{i}\mathbf{e}^{i}+v_{i}\mathbf{de}^{i} \\
&=&\left( \mathbf{d}v_{k}-v_{i}\mathbf{\omega }_{k}^{\quad i}\right) \mathbf{%
e}^{k}
\end{eqnarray*}
Similar use of the product rule gives the covariant derivative of higher
rank tensors. This local $SO\left( 3\right) $-covariant derivative of forms
in an orthonormal basis is equivalent to a general coordinate covariant
derivative when expressed in terms of a coordinate basis. We see this as
follows.

Rewriting eq.(\ref{Solder form}) in the form 
\[
\mathbf{de}^{i}+\mathbf{e}^{k}\mathbf{\omega }_{\quad k}^{i}=0 
\]
we expand in an arbitrary coordinate basis, to find 
\[
\mathbf{d}x^{\alpha }\wedge \mathbf{d}x^{\beta }\left( \partial _{\alpha }%
e_{\beta }^{\quad i}+e_{\alpha }^{\quad k}\omega _{\quad
k\beta }^{i}\right) =0 
\]
The term in parentheses must therefore be symmetric: 
\[
\partial _{\alpha }e_{\beta }^{\quad i}+e_{\alpha }^{\quad
k}\omega _{k\beta }^{i}\equiv \Gamma _{\alpha \beta }^{i}=\Gamma _{\beta
\alpha }^{i} 
\]
Writing 
\[
\Gamma _{\beta \alpha }^{i}=e_{\mu }^{\quad i}\Gamma _{\beta \alpha
}^{\mu } 
\]
we define the covariant constancy of the basis coefficients, 
\begin{equation}
D_{\alpha }e_{\beta }^{\quad i}\equiv \partial _{\alpha }e_{\beta }^{\quad i}+e_{\alpha }^{\quad k}\omega _{k\beta }^{i}-%
e_{\mu }^{\quad i}\Gamma _{\beta \alpha }^{\mu }=0
\label{Constancy of  e}
\end{equation}
Eq.(\ref{Constancy of e}) relates the $SO(3)$-covariant spin connection for
orthonormal frames to the Christoffel connection for general coordinate
transformations. Since the covariant derivative of the orthogonal metric $%
\eta =diag(1,1,1)$ is zero, 
\begin{eqnarray*}
D_{\alpha }\eta _{ab} &=&\partial
_{\alpha }\eta _{ab}-\eta _{cb}\mathbf{\omega }_{a}^{c}-\eta _{ac}\mathbf{%
\omega }_{b}^{c} \\
&=&-\eta _{cb}\mathbf{\omega }_{a}^{c}-\eta _{ac}\mathbf{\omega }_{b}^{c} \\
&=&0
\end{eqnarray*}
where the last step follows by the antisymmetry of the $SO(3)$ connection,
we have covariant constancy of the metric: 
\[
D_{\alpha }g_{\mu \nu }=D_{\alpha }\left( \eta _{ab}e_{\mu }^{\quad
a}e_{\nu }^{\quad b}\right) =0 
\]
This is inverted in the usual way to give the Christoffel connection for $%
SO(3),$%
\begin{equation}
\Gamma _{\mu \nu }^{\alpha }=\frac{1}{2}g^{\alpha \beta }\left( g_{\beta \mu
,\nu }+g_{\beta \nu ,\mu }-g_{\mu \nu ,\beta }\right)  \label{Christoffel}
\end{equation}
The Christoffel connection may also be found directly from eq.(\ref
{Constancy of e}) using $g_{\mu \nu }=\eta _{ab}e_{\mu }^{\quad a}%
e_{\nu }^{\quad b}$. There is little practical difference between
the ability to perform local rotations on an orthonormal frame field, and
the ability to perform arbitrary transformations of coordinates. It is just
a matter of putting the emphasis on the coordinates or on the basis vectors
(see \cite{Isham}). It is this equivalence that makes the $SO(3)$ gauge
theory equivalent to the use of ``generalized coordinates'' in Lagrangian
mechanics.

Since Newtonian 3-space is Euclidean and we have not generalized to curved
spaces, the metric is always just a diffeomorphism away from orthonormal,
that is, 
\begin{eqnarray}
e_{\alpha }^{\quad a} &=&J_{\alpha }^{\quad a}=\frac{\partial y^{a}}{%
\partial x^{\alpha }}  \nonumber \\
g_{\alpha \beta } &=&\eta _{ab}\mathbf{e}_{\alpha }^{\quad a}\mathbf{e}%
_{\beta }^{\quad b}=\eta _{ab}\frac{\partial y^{a}}{\partial x^{\alpha }}%
\frac{\partial y^{b}}{\partial x^{\beta }}  \label{Euclidean metric}
\end{eqnarray}
and the connection takes the simple form 
\begin{equation}
\Gamma _{\mu \nu }^{\alpha }=-\frac{\partial x^{\alpha }}{\partial y^{a}}%
\frac{\partial ^{2}y^{a}}{\partial x^{\mu }\partial x^{\nu }}
\label{Euclidean connection}
\end{equation}
which has, of course, vanishing curvature. Notice that $\left( \mathbf{e}%
^{a},\mathbf{\omega }_{\quad c}^{a}\right) $ or equivalently, $\left( g_{\alpha \beta
},\Gamma _{\mu \nu }^{\alpha }\right) $, here describe a much larger class of
coordinate transformations than the global conformal connection $\Lambda
_{jk}^{i}$ of Sec. 4. The connection of eq.(\ref{Euclidean connection})
gives a derivative which is covariant for \textit{any} coordinate
transformation.

This completes our description of Euclidean 3-space in general $SO(3)$
frames or general coordinates. We now generalize Newton's second law to be
consistent with the enhanced symmetry.

\subsection{Generally covariant form of Newton's law}

The generalization of Newton's second law to a locally $SO(3)$ covariant
form of mechanics is now immediate. We need only replace the time derivative
by a directional covariant derivative, 
\[
F^{i}=v^{k}D_{k}\left( mv^{i}\right) 
\]
where 
\begin{equation}
D_{k}v^{i}\equiv \partial _{k}v^{i}+v^{j}\Gamma _{jk}^{i}
\label{Gen Coord Cov derivative}
\end{equation}
and $\Gamma _{jk}^{i}$ is given by eq.(\ref{Euclidean connection}). This is
the principal result our the $SO(3)$ gauging.

If $F^{i}$ is curl free, then it may be written as the contravariant form of
the gradient of a potential 
\[
F^{i}=-g^{ij}\frac{\partial \phi }{\partial x^{j}} 
\]
The covariant form of Newton's law then follows as the extremum of the
action 
\[
S=\int dt\left( \frac{m}{2}g_{mn}\frac{dx^{m}}{dt}\frac{dx^{n}}{dt}-\phi
\right) 
\]
The integrand is identical to the usual Lagrangian for classical mechanics,
so the $ISO(3)$ gauge theory has led us to Lagrangian mechanics. Indeed, it
is straightforward to check that the terms of the Euler-Lagrange equation
involving the kinetic energy 
\[
T=\frac{m}{2}g_{mn}\frac{dx^{m}}{dt}\frac{dx^{n}}{dt} 
\]
where $g_{mn}$ is given by eq.(\ref{Euclidean metric}), combine to give
precisely the covariant acceleration: 
\begin{equation}
\frac{d}{dt}\frac{\partial T}{\partial v^{i}}-\frac{\partial T}{\partial
x^{i}}=mv^{m}D_{m}v^{i}
\end{equation}
with $D_{m}v^{i}$ as in eq.(\ref{Gen Coord Cov derivative}) and $\Gamma
_{jk}^{i}$ as in eq.(\ref{Christoffel}). This equivalence is the central
result of this section. This result is expected, since Lagrangian mechanics
was formulated in order to allow ``generalized coordinates'', i.e., general
coordinate transformations, but the equivalence is rarely pointed out.

We treat multiple particles in the usual way. Suppose we have Newton's
second law for each of $N$ variables. Then we have the system 
\[
F_{A}^{i}=m_{A}a_{A}^{i} 
\]
for $A=1,2,\ldots ,N.$ The symmetry, however, applies in the same way to
each particle's coordinates, so the symmetry of the full system is still
described by $ISO(3)$. When we gauge $ISO(3)$ we therefore still have a
coordinate-invariant formulation of Euclidean 3-space with connection and
metric as given in eqs.(\ref{Euclidean metric}) and (\ref{Euclidean
connection}). The only change comes in the action, which we now write as a
sum over all particles: 
\[
S=\sum_{A}\int dt\left( \frac{m_{A}}{2}g_{mn}\left( x_{A}\right) \frac{%
dx_{A}^{m}}{dt}\frac{dx_{A}^{n}}{dt}-\phi \left( x_{A}\right) \right) 
\]
Notice how the metric suffices for all $N$ particles because it is a
function of position. Each term in the sum causes the metric to be evaluated
at a different point.

We have therefore shown that the locally $SO(3)$-covariant gauge theory of
Newton's second law is Lagrangian mechanics.

We now repeat the procedure for the full conformal symmetry associated with
Newtonian measurement theory.

\section{A conformal gauge theory of Newtonian mechanics}

Now we gauge the full $O(4,1)$ symmetry of our globally conformal form of
Newton's law. The Lie algebra of the conformal group (see Appendix 5) is:

\begin{eqnarray}
\left[ M_{\quad b}^{a},M_{\quad d}^{c}\right] &=&\delta _{b}^{c}M_{\quad
d}^{a}-\eta ^{ca}\eta _{be}M_{\quad d}^{e}-\eta _{bd}M^{ac}+\delta
_{d}^{a}M_{b}^{\quad c}  \nonumber \\
\left[ M_{\quad b}^{a},P_{c}\right] &=&\eta _{bc}\eta ^{ae}P_{e}-\delta
_{c}^{a}P_{b}  \nonumber \\
\left[ M_{\quad b}^{a},K^{c}\right] &=&\delta _{b}^{c}K^{a}-\eta ^{ca}\eta
_{be}K^{e}  \nonumber \\
\left[ P_{b},K_{d}\right] &=&-\eta _{be}M_{\quad d}^{e}-\eta _{bd}D 
\nonumber \\
\left[ D,P_{a}\right] &=&-P_{a}  \nonumber \\
\left[ D,K^{a}\right] &=&K^{a}  \label{Lie Algebra}
\end{eqnarray}
where $M_{\quad b}^{a},P_{a},K_{a}$ and $D$ generate rotations,
translations, special conformal transformations and dilatations,
respectively.

As before, we write the Lie algebra in terms of the dual basis of 1-forms,
setting 
\begin{eqnarray*}
\left\langle M_{\quad b}^{a},\mathbf{\omega }_{\quad d}^{c}\right\rangle
&=&\delta _{b}^{c}\delta _{d}^{a}-\eta ^{ca}\eta _{be} \\
\left\langle P_{b},\mathbf{e}^{a}\right\rangle &=&\delta _{b}^{a} \\
\left\langle K^{a},\mathbf{f}_{b}\right\rangle &=&\delta _{b}^{a} \\
\left\langle D,\mathbf{W}\right\rangle &=&1
\end{eqnarray*}
The Maurer-Cartan structure equations are therefore 
\begin{eqnarray}
\mathbf{d\omega }_{\quad b}^{a} &=&\mathbf{\omega }_{\quad b}^{c}\mathbf{\
\omega }_{\quad c}^{a}+\mathbf{f}_{b}\mathbf{e}^{a}-\eta ^{ac}\eta _{bd}%
\mathbf{f}_{c}\mathbf{e}^{d}  \label{Spin connection MC eq} \\
\mathbf{de}^{a} &=&\mathbf{e}^{c}\mathbf{\omega }_{\quad c}^{a}+\mathbf{We}%
^{a}  \label{Solder form MC eq} \\
\mathbf{df}_{a} &=&\mathbf{\omega }_{\quad a}^{c}\mathbf{f}_{c}+\mathbf{f}%
_{a}\mathbf{W}  \label{Co-solder form} \\
\mathbf{dW} &=&\mathbf{e}^{a}\mathbf{f}_{a}  \label{Weyl vector}
\end{eqnarray}
So far, these structure equations look the same regardless of how the group
is gauged. However, there are different ways to proceed from here because
there is more than one sensible subgroup. In principle, we may take the
quotient of the conformal group by any subgroup, as long as that subgroup
contains no normal subgroup of the conformal group. However, we certainly
want the final result to permit local rotations and local dilatations.
Looking at the Lie algebra, we see only three subgroups\footnote{%
We consider only rotationally and dilatationally covariant subgroups, which
restricts consideration to subsets of $\{M_{\quad b}^{a},P_{a},K_{a},D\}$
and not, for example, collections such as $\{P_{2},K_{2},D\}.$} satisfying
this condition, namely, those generated by one of the following three sets
of generators 
\begin{eqnarray*}
&&\left\{ M_{\quad b}^{a},P_{a},D\right\} \\
&&\left\{ M_{\quad b}^{a},K_{a},D\right\} \\
&&\left\{ M_{\quad b}^{a},D\right\}
\end{eqnarray*}
The first two generate isomorphic subgroups, so there are really only two
independent choices, $\left\{ M_{\quad b}^{a},K_{a},D\right\} $ and $\left\{
M_{\quad b}^{a},D\right\} $. The most natural choice is the first because it
results once again in a gauge theory of a 3-dim Euclidean space. However, it
leads only to a conformally flat 3-geometry with no new features. The final
possibility, $\left\{ M_{\quad b}^{a},D\right\} ,$ is called biconformal
gauging. It turns out to be interesting.

Therefore, we perform the biconformal gauging, choosing the homogeneous Weyl
group generated by $\left\{ M_{\quad b}^{a},D\right\} $ for the local
symmetry. This means that the forms $\mathbf{e}^{a}$ and $\mathbf{f}_{a}$
are independent, spanning a 6-dimensional sub-manifold of the conformal
group manifold.

The solution of the structure equations (see \cite{NewConfGauging}), eqs.(%
\ref{Spin connection MC eq}-\ref{Weyl vector}) may be put in the form:

\begin{eqnarray*}
\mathbf{\omega }_{\quad b}^{a} &=&\left( \delta _{d}^{a}\delta _{b}^{c}-\eta
^{ac}\eta _{db}\right) y_{c}\mathbf{\mathbf{d}}x^{d} \\
\mathbf{W} &=&-y_{a}\mathbf{\mathbf{d}}x^{a} \\
\mathbf{e}^{a} &=&\mathbf{d}x^{a} \\
\mathbf{f}_{a} &=&\mathbf{d}y_{a}-\left( y_{a}y_{b}-\frac{1}{2}y^{2}\eta
_{ab}\right) \mathbf{d}x^{b}
\end{eqnarray*}
Notice that if we hold $y_{a}$ constant, these forms are defined on a 3-dim
space with orthonormal basis $\mathbf{e}^{a}=\mathbf{d}x^{a}.$ Since this is
also a coordinate basis, this subspace is Euclidean, and we identify it with
the original configuration space.

We can see that $\mathbf{e}^{a}\mathbf{f}_{a}$ is a symplectic form because $%
\mathbf{e}^{a}$ and $\mathbf{f}_{a}$ are independent, making this 2-form
non-degenerate, while the structure equation, eq.(\ref{Weyl vector}), 
\begin{equation}
\mathbf{dW}=\mathbf{e}^{a}\mathbf{f}_{a}
\end{equation}
shows that $\mathbf{e}^{a}\mathbf{f}_{a}$ is closed, $\mathbf{d}\left( 
\mathbf{e}^{a}\mathbf{f}_{a}\right) =\mathbf{d}^{2}\mathbf{W}=0.$ This is
also evident from the solution, where 
\begin{eqnarray*}
\mathbf{e}^{a}\mathbf{f}_{a} &=&\mathbf{d}x^{a}\left( \mathbf{d}y_{a}-\left(
y_{a}y_{b}-\frac{1}{2}y^{2}\eta _{ab}\right) \mathbf{d}x^{b}\right) \\
&=&\mathbf{d}x^{a}\mathbf{d}y_{a}
\end{eqnarray*}
is in canonical form. Because of this symplectic form we are justified in
identifying the solution as a relative of phase space.

Since we are in some 6-dimensional space, we cannot simply write Newton's
law as before. Moreover, with the interpretation as a relative of phase
space, we do not expect physical paths to be geodesics. We start then with
an action. Noting that the geometry contains a new one-form, the Weyl
vector, it is reasonable to examine what paths are determined by its
extremals. Therefore, we consider the action 
\[
S_{0}=\int \mathbf{W} 
\]
We add a function to make it interesting (note that the relativistic case
adds this function automatically). Then 
\begin{eqnarray*}
S &=&\int \left( \mathbf{W}+f\mathbf{d}t\right) \\
&=&\int \left( -y_{m}dx^{m}+fdt\right) \\
&=&-\int \left( y_{m}\frac{dx^{m}}{dt}-f\right) dt
\end{eqnarray*}
Since the function now depends on six variables, $x^{m},y_{m}$ the variation
gives 
\begin{eqnarray*}
0 &=&\delta S \\
&=&-\int \left( \delta y_{m}\frac{dx^{m}}{dt}+y_{m}\frac{d\delta x^{m}}{dt}-%
\frac{\partial f}{\partial x^{m}}\delta x^{m}-\frac{\partial f}{\partial
y_{m}}\delta y_{m}\right) dt \\
&=&-\int \left( \left( \frac{dx^{m}}{dt}-\frac{\partial f}{\partial y_{m}}%
\right) \delta y_{m}+\left( -\frac{dy_{m}}{dt}-\frac{\partial f}{\partial
x^{m}}\right) \delta x^{m}\right) dt
\end{eqnarray*}
so that 
\begin{eqnarray*}
\frac{dx^{m}}{dt} &=&\frac{\partial f}{\partial y_{m}} \\
\frac{dy_{m}}{dt} &=&-\frac{\partial f}{\partial x^{m}}
\end{eqnarray*}
We recognize these immediately as Hamilton's equations, if we identify the
function $f$ with the Hamiltonian.

As expected, the symmetry of these equations includes local rotations and
local dilatations, but in fact is larger since, as we know, local symplectic
transformations preserve Hamilton's equations.

Hamilton's principal function follows by evaluating the action along the
classical paths of motion. This function is well-defined because the
Lagrangian one-form is curl free on these paths: 
\begin{eqnarray*}
\mathcal{L} &=&y_{m}\mathbf{d}x^{m}-f\mathbf{d}t \\
\mathbf{d}\mathcal{L} &=&\mathbf{d}y_{m}\wedge \mathbf{d}x^{m}-\mathbf{d}%
f\wedge \mathbf{d}t \\
&=&\mathbf{d}y_{m}\wedge \mathbf{d}x^{m}-\frac{\partial f}{\partial y_{m}}%
\mathbf{d}y_{m}\wedge \mathbf{d}t-\frac{\partial f}{\partial x^{m}}\mathbf{d}%
x^{m}\wedge \mathbf{d}t \\
&=&\left( \mathbf{d}y_{m}+\frac{\partial f}{\partial x^{m}}\mathbf{d}%
t\right) \wedge \left( \mathbf{d}x^{m}-\frac{\partial f}{\partial y_{m}}%
\mathbf{d}t\right) \\
&=&0
\end{eqnarray*}
where the last step follows by imposing the equations of motion. Thus, the
integral of 
\[
\mathcal{L}=y_{m}\mathbf{d}x^{m}-f\mathbf{d}t 
\]
along solutions to Hamilton's equations is independent of path, and 
\[
\mathcal{S}\left( x\right) =\int_{x_{0}}^{x}\mathcal{L} 
\]
unambiguously gives Hamilton's principal function.

\subsection{Multiple particles}

Multiple particles in the conformal gauge theory are handled in a way
similar to the Newtonian case. This is an important difference from the
usual treatment of Hamiltonian dynamics, where the size of the phase space
depends on the number of particles. Here, there is a single conformal
connection on the full conformal group manifold, and gauging still gives a
6-dimensional space with a symplectic form. The variational principal
becomes a sum over all particles, 
\begin{eqnarray*}
S &=&\sum_{A=1}^{N}\int \left( \mathbf{W}\left( x_{A}^{m},y_{m}^{A}\right)
+f\left( x_{A}^{m},y_{m}^{A}\right) dt\right) \\
&=&-\sum_{A=1}^{N}\int \left( y_{m}^{A}\frac{dx_{A}^{m}}{dt}-f\left(
x_{A}^{m},y_{m}^{A}\right) dt\right)
\end{eqnarray*}
Variation with respect to all $6N$ particle coordinates then yields the
usual set of Hamilton's equations. If we regard the Lagrangian 1-form as depending independently on all $N$ particle positions,
\[
\mathbf{L}= -\sum_{A=1}^{N} \left( y_{m}^{A}\frac{dx_{A}^{m}}{dt}-f\left(
x_{A}^{m},y_{m}^{A}\right) dt\right)
\]
then $\mathbf{dL}=0$ as before and Hamilton's principal function is again well-defined.

The situation here is to be contrasted with the usual $N$-particle phase
space, which is of course $6N$-dimensional. Instead we have a $6$%
-dimensional symplectic space in which all $N$ particles move. The idea is
to regard biconformal spaces as fundamental in the same sense as
configuration spaces, rather than derived from dynamics the way that phase
spaces are. This means that in principle, dynamical systems could depend on
position and momentum variables independently. Classical solutions, however,
have been shown to separate neatly into a pair of 3-dimensional submanifolds
with the usual properties of configuration space and momentum space.
Similarly, relativistic solutions with curvature separate into a pair of
4-dim manifolds with the properties of spacetime and energy-momentum space,
with the Einstein equation holding on the spacetime submanifold. These
observations suggest that the symplectic structure encountered in dynamical
systems might actually have kinematic (i.e., symmetry based) origins.

\section{Is size change measurable?}

While we won't systematically introduce curvature, there is one important
consequence of dilatational curvature that we must examine. A full
examination of the field equations for curved biconformal space (\cite
{NewConfGauging},\cite{WW} ) shows that the dilatational curvature is
proportional (but not equal) to the curl of the Weyl vector. When this
curvature is nonzero, the relative sizes of physical objects may change.
Specifically, suppose two initially identical objects move along paths
forming the boundary to a surface. If the integral of the dilatational
curvature over that surface does not vanish the two objects will no longer
have identical sizes. This result is inconsistent with macroscopic physics.
However, we now show that the result never occurs classically. A similar
result has been shown for Weyl geometries \cite{QuantMeasandGeom}.

If we fix a gauge, the change in any length dimension, $l,$ along any path, $%
C,$ is given by the integral of the Weyl vector along that path: 
\begin{eqnarray*}
dl &=&lW_{i}dx^{i} \\
l &=&l_{0}\exp \left( \int_{C}W_{i}dx^{i}\right)
\end{eqnarray*}
It is this integral that we want to evaluate for the special case of
classical paths. Notice that this factor is gauge dependent, but if we
compare two lengths which follow different paths with common endpoints, the
ratio of their lengths changes in a gauge independent way: 
\[
\frac{l_{1}}{l_{2}}=\frac{l_{10}}{l_{20}}\exp \left(
\oint_{C_{1}-C_{2}}W_{i}dx^{i}\right) 
\]
This dilatation invariant result represents measurable relative size
change.

We now show that such measurable size changes never occur classically. From
the expression for the action we have 
\[
\int W_{i}dx^{i}=\int W_{i}\frac{dx^{i}}{dt}dt=S(x)-\int^{x}fdt 
\]
where we write the action as a function of $x$ because we are evaluating
only along classical paths, where $S$ becomes Hamilton's principle function.
This is precisely what we need. The right side of this expression is a
function; its value is independent of the path of integration. Therefore,
the integral of the Weyl vector along every classical path may be removed
(all of them at once) by the gauge transformation 
\[
e^{-S(x)+\int fdt} 
\]
Then in the new gauge, 
\begin{eqnarray*}
\int W_{i}^{\prime }dx^{i} &=&\int \left( W_{i}-\partial _{i}S(x)+\partial
_{i}\int fdt\right) dx^{i} \\
&=&\int W_{i}dx^{i}-S(x)+\int fdt \\
&=&0
\end{eqnarray*}
regardless of the (classical) path of integration. Therefore, no classical
objects ever display measurable length change.

\section{Conclusions}

We have shown the following

\begin{enumerate}
\item  The $SO(3)$ gauge theory of Newton's second law is Lagrangian
mechanics

\item  The $SO(4,1)$ gauge theory of Newton's second law gives Hamilton's
equations of motion on a fundamental 6-dim symplectic space.
\end{enumerate}

These results provide a new unification of classical mechanics using the
tools of gauge theory.

We note several further insights.

First, by identifying the symmetries of a theory's dynamical law from the
symmetry of its measurement theory, we gain new insight into the meaning of
gauge theory. Generally speaking, dynamical laws will have global symmetries
while the inner products required for measurement will have local
symmetries. Gauging may be viewed as enlarging the symmetry of the dynamical
law to match the symmetry of measurement, thereby maintaining closer contact
with what is, in fact, measurable.

Second, we strengthen our confidence and understanding of the interpretation of relativistic biconformal spaces
as relatives of phase space. The fact that the same gauging applied to
classical physics yields the well-known and powerful formalism of
Hamiltonian dynamics suggests that the higher symmetry of biconformal
gravity theories may in time lead to new insights or more powerful solution
techniques.

Finally, it is possible that the difference between the 6-dimensional
symplectic space of $SO(4,1)$ gauge theory and $2N$-dimensional phase space
of Hamiltonian dynamics represents a deep insight. Like Hamiltonian
dynamics, quantum mechanics requires both position and momentum variables
for its formulation -- without both, the theory makes no sense. If we take
this seriously, perhaps we should look closely at a higher dimensional
symplectic space as the fundamental arena for physics. Rather than regarding 
$2N$-dim phase space as a convenience for calculation, perhaps there is a
6-dim (or, relativistically, 8-dim) space upon which we move and make our
measurements. If this conjecture is correct, it will be interesting to see
the form taken by quantum mechanics or quantum field theory when formulated
on a biconformal manifold.

The proof of Sec. 8 is encouraging in this regard, for not only do classical paths show no dilatation, but a converse statement holds as well: non-classical paths generically do show dilatation. Since quantum systems may be regarded as sampling all paths (as in a path integral), it may be possible to regard quantum non-integrability of phases as related to non-integrable size change. There is a good reason to think that this correspondence occurs: biconformal spaces have a metric structure (consistent with classical collisions \cite{NewConfGauging}) which projects separately onto its position and momentum subspaces. However, while the configuration space metric and momentum space metric are necessarily identical, the biconformal projections have opposite signs. This difference is reconciled if we identify the biconformal coordinate $y_{k}$ with $i$ times the momentum. This identification does not alter the classical results, but it changes the dilatations to phase transformations when $y_{k}$ is replaced by $ip_{k}.$ If this is the case, then the evolution of sizes in biconformal spaces, when expressed in the usual classical variables, gives unitary evolution just as in quantum physics. The picture here is much like the familiar treatment of quantum systems as thermodynamic systems by replacing time by a complex temperature parameter, except it is now the energy-momentum vector that is replaced by a complex coordinate in a higher dimensional space.  A full examination of these questions takes us too far afield to pursue here, but they are under current investigation.

\bigskip

\noindent {\large Acknowledgments}

The author wishes to thank Stefan Hollands for a question prompting parts of
this investigation, and Lara B. Anderson, Clayton Call, Charles Torre, and
David Peak for interesting and useful discussions.

\pagebreak

\pagebreak

{\Large \noindent }{\large Appendices}

\bigskip

\noindent {\large Appendix 1: Point transformations of Newton's second law}

Here we derive the point transformations leaving the second law invariant,
assuming the force to transform as a vector.

Consider a general coordinate transformation in which we replace the
Cartesian coordinates, $x^{i},$ as well as the time parameter, by 
\begin{eqnarray*}
q^{i} &=&q^{i}\left( \mathbf{x},t\right) \\
\tau &=&\tau \left( \mathbf{x},t\right)
\end{eqnarray*}
We have four functions, each of four variables. This functions must be
invertible, so we may also write 
\begin{eqnarray*}
x^{i} &=&x^{i}\left( \mathbf{q},\tau \right) \\
t &=&t\left( \mathbf{q},\tau \right)
\end{eqnarray*}
The limitation on covariance comes from the acceleration. First, the
velocity is given by 
\begin{eqnarray*}
v^{i} &=&\frac{dx^{i}\left( \mathbf{q},\tau \right) }{dt} \\
&=&\frac{d\tau }{dt}\left( \frac{\partial x^{i}}{\partial q^{j}}\frac{dq^{j}%
}{d\tau }+\frac{\partial x^{i}}{\partial \tau }\right)
\end{eqnarray*}
where we use the usual summation convention on repeated indices, e.g., 
\[
\sum_{j=1}^{3}\frac{\partial x^{i}}{\partial q^{j}}\frac{dq^{j}}{d\tau }=%
\frac{\partial x^{i}}{\partial q^{j}}\frac{dq^{j}}{d\tau } 
\]
The acceleration is 
\begin{eqnarray*}
a^{i} &=&\frac{dv^{i}\left( \mathbf{q},\tau \right) }{dt} \\
&=&\frac{d\tau }{dt}\frac{d}{d\tau }\left( \frac{d\tau }{dt}\left( \frac{%
\partial x^{i}}{\partial q^{j}}\frac{dq^{j}}{d\tau }+\frac{\partial x^{i}}{%
\partial \tau }\right) \right) \\
&=&\left( \frac{d\tau }{dt}\right) ^{2}\left( \frac{\partial x^{i}}{\partial
q^{j}}\frac{d^{2}q^{j}}{d\tau ^{2}}+\frac{\partial x^{i}}{\partial \tau }%
\right) +\frac{d\tau }{dt}\frac{d^{2}\tau }{dt^{2}}\left( \frac{\partial
x^{i}}{\partial q^{j}}\frac{dq^{j}}{d\tau }+\frac{\partial x^{i}}{\partial
\tau }\right) \\
&&+\left( \frac{d\tau }{dt}\right) ^{2}\frac{dq^{k}}{d\tau }\left( \frac{%
\partial ^{2}x^{i}}{\partial q^{k}\partial q^{j}}\frac{dq^{j}}{d\tau }+\frac{%
\partial ^{2}x^{i}}{\partial q^{k}\partial \tau }\right) \\
&&+\left( \frac{d\tau }{dt}\right) ^{2}\left( \frac{\partial ^{2}x^{i}}{%
\partial \tau \partial q^{j}}\frac{dq^{j}}{d\tau }+\frac{\partial ^{2}x^{i}}{%
\partial \tau ^{2}}\right)
\end{eqnarray*}
The first term is proportional to the acceleration of $q^{i}$, but the
remaining terms are not. Since we assume that force is a vector, it changes
according to: 
\begin{equation}
F^{i}\left( \mathbf{x},t\right) =\frac{\partial x^{i}}{\partial q^{j}}%
F^{j}\left( \mathbf{q},\tau \right)  \label{Vector trans law}
\end{equation}
where $\frac{\partial x^{i}}{\partial q^{j}}$ is the Jacobian matrix of the
coordinate transformation. Substituting into the equation of motion, we have 
\begin{eqnarray}
\frac{1}{m}\frac{\partial x^{i}}{\partial q^{j}}F^{j}\left( \mathbf{q},\tau
\right) &=&\left( \frac{d\tau }{dt}\right) ^{2}\left( \frac{\partial x^{i}}{%
\partial q^{j}}\frac{d^{2}q^{j}}{d\tau ^{2}}+\frac{\partial x^{i}}{\partial
\tau }\right)  \nonumber \\
&&+\frac{d\tau }{dt}\frac{d^{2}\tau }{dt^{2}}\left( \frac{\partial x^{i}}{%
\partial q^{j}}\frac{dq^{j}}{d\tau }+\frac{\partial x^{i}}{\partial \tau }%
\right)  \nonumber \\
&&+\left( \frac{d\tau }{dt}\right) ^{2}\frac{dq^{k}}{d\tau }\left( \frac{%
\partial ^{2}x^{i}}{\partial q^{k}\partial q^{j}}\frac{dq^{j}}{d\tau }+\frac{%
\partial ^{2}x^{i}}{\partial q^{k}\partial \tau }\right)  \nonumber \\
&&+\left( \frac{d\tau }{dt}\right) ^{2}\left( \frac{\partial ^{2}x^{i}}{%
\partial \tau \partial q^{j}}\frac{dq^{j}}{d\tau }+\frac{\partial ^{2}x^{i}}{%
\partial \tau ^{2}}\right)   \label{Transf Newton}
\end{eqnarray}
Multiplying by the inverse to the Jacobian matrix, $\left( \frac{\partial
q^{m}}{\partial x^{i}}\right) ,$ eq.(\ref{Transf Newton}) becomes 
\begin{eqnarray}
\frac{1}{m}F^{m}\left( \mathbf{q},\tau \right) &=&\left( \frac{d\tau }{dt}%
\right) ^{2}\left( \frac{d^{2}q^{m}}{d\tau ^{2}}+\frac{\partial q^{m}}{%
\partial x^{i}}\frac{\partial x^{i}}{\partial \tau }\right)  \nonumber \\
&&+\frac{d\tau }{dt}\frac{d^{2}\tau }{dt^{2}}\left( \frac{dq^{m}}{d\tau }+%
\frac{\partial q^{m}}{\partial x^{i}}\frac{\partial x^{i}}{\partial \tau }%
\right)  \nonumber \\
&&+\left( \frac{d\tau }{dt}\right) ^{2}\left( \frac{\partial q^{m}}{\partial
x^{i}}\frac{\partial ^{2}x^{i}}{\partial q^{k}\partial q^{j}}\frac{dq^{k}}{%
d\tau }\frac{dq^{j}}{d\tau }+\frac{dq^{k}}{d\tau }\frac{\partial q^{m}}{%
\partial x^{i}}\frac{\partial ^{2}x^{i}}{\partial q^{k}\partial \tau }\right)
 \nonumber \\
&&+\left( \frac{d\tau }{dt}\right) ^{2}\left( \frac{\partial q^{m}}{\partial
x^{i}}\frac{\partial ^{2}x^{i}}{\partial \tau \partial q^{j}}\frac{dq^{j}}{%
d\tau }+\frac{\partial q^{m}}{\partial x^{i}}\frac{\partial ^{2}x^{i}}{%
\partial \tau ^{2}}\right)
\end{eqnarray}
Therefore, Newton's second law holds in the new coordinate system 
\[
F^{m}\left( \mathbf{q},\tau \right) =m\frac{d^{2}q^{m}}{d\tau ^{2}} 
\]
if and only if: 
\begin{eqnarray}
1 &=&\left( \frac{d\tau }{dt}\right) ^{2} \\
0 &=&\frac{\partial q^{m}}{\partial x^{i}}\frac{\partial x^{i}}{\partial
\tau }  \\
0 &=&\frac{d\tau }{dt}\frac{d^{2}\tau }{dt^{2}}\left( \frac{dq^{m}}{d\tau }+%
\frac{\partial q^{m}}{\partial x^{i}}\frac{\partial x^{i}}{\partial \tau }%
\right)  \nonumber  \\
&&+\left( \frac{d\tau }{dt}\right) ^{2}\left( \frac{\partial q^{m}}{\partial
x^{i}}\frac{\partial ^{2}x^{i}}{\partial q^{k}\partial q^{j}}\frac{dq^{k}}{%
d\tau }\frac{dq^{j}}{d\tau }+\frac{dq^{k}}{d\tau }\frac{\partial q^{m}}{%
\partial x^{i}}\frac{\partial ^{2}x^{i}}{\partial q^{k}\partial \tau }\right)
 \nonumber \\
&&+\left( \frac{d\tau }{dt}\right) ^{2}\left( \frac{\partial q^{m}}{\partial
x^{i}}\frac{\partial ^{2}x^{i}}{\partial \tau \partial q^{j}}\frac{dq^{j}}{%
d\tau }+\frac{\partial q^{m}}{\partial x^{i}}\frac{\partial ^{2}x^{i}}{%
\partial \tau ^{2}}\right)
\end{eqnarray}
Therefore, we have 
\[
\tau =t+t_{0} 
\]
together with the possibility of time reversal, 
\[
\tau =-t+t_{0} 
\]
for the time parameter. Using the first two equations to simplify the third
(including $\frac{d^{2}\tau }{dt^{2}}=0),$%
\begin{eqnarray}
0 &=&\left( \frac{\partial q^{m}}{\partial x^{i}}\frac{\partial ^{2}x^{i}}{%
\partial q^{k}\partial q^{j}}\right) \frac{dq^{k}}{d\tau }\frac{dq^{j}}{%
d\tau }  \nonumber \\
&&+2\frac{\partial q^{m}}{\partial x^{i}}\frac{\partial ^{2}x^{i}}{\partial
q^{k}\partial \tau }\frac{dq^{k}}{d\tau }  \nonumber \\
&&+\frac{\partial ^{2}x^{i}}{\partial \tau ^{2}}\frac{\partial q^{m}}{%
\partial x^{i}}
\end{eqnarray}
Now, since the components of the velocity, $\frac{dq^{k}}{d\tau },$ are
independent we get three equations, 
\begin{eqnarray}
0 &=&\frac{\partial q^{m}}{\partial x^{i}}\frac{\partial ^{2}x^{i}}{\partial
q^{k}\partial q^{j}} \\
0 &=&\frac{\partial q^{m}}{\partial x^{i}}\frac{\partial ^{2}x^{i}}{\partial
q^{k}\partial \tau } \\
0 &=&\frac{\partial q^{m}}{\partial x^{i}}\frac{\partial ^{2}x^{i}}{\partial
\tau ^{2}}
\end{eqnarray}
Since the Jacobian matrix is invertible, these reduce to 
\begin{eqnarray}
0 &=&\frac{\partial ^{2}x^{i}}{\partial q^{k}\partial q^{j}} \\
0 &=&\frac{\partial ^{2}x^{i}}{\partial q^{k}\partial \tau } \\
0 &=&\frac{\partial ^{2}x^{i}}{\partial \tau ^{2}}
\end{eqnarray}
and integrating, 
\begin{eqnarray}
0 &=&\frac{\partial ^{2}x^{i}}{\partial \tau ^{2}}\Rightarrow
x^{i}=x_{0}^{i}\left( q^{m}\right) +v_{0}^{i}\left( q^{m}\right) \tau \\
0 &=&\frac{\partial ^{2}x^{i}}{\partial q^{k}\partial \tau }\Rightarrow 0=%
\frac{\partial v_{0}^{i}}{\partial q^{k}}\Rightarrow v_{0}^{i}=const.
\end{eqnarray}

The remaining equation implies that the Jacobian matrix is constant, 
\begin{equation}
\frac{\partial x^{m}}{\partial q^{j}}=\frac{\partial x_{0}^{m}}{\partial
q^{j}}=J_{\;\;j}^{m}=const.
\end{equation}
Integrating, the coordinates must be related by a constant, inhomogeneous,
general linear transformation, 
\begin{eqnarray}
x^{m} &=&J_{\;\;j}^{m}q^{j}+v_{0}^{i}\tau +x_{0}^{m} \\
t &=&\tau +\tau _{0}
\end{eqnarray}
together with a shift and possible time reversal of $t.$

We get a $16$-parameter family of coordinate systems: nine for the
independent components of the nondegenerate $3\times 3$ matrix $J,$ three
for the boosts $v_{0}^{i},$ three more for the arbitrary translation, $%
x_{0}^{m},$ and a single time translation.

Notice that the transformation includes the possibility of an arbitrary
scale factor, $e^{-2\lambda }=\left| \det \left( J_{\quad n}^{m}\right)
\right| .$

\pagebreak

\noindent {\large Appendix 2: Special conformal transformations}

In three dimensions, there are ten independent transformations preserving
the inner product (or the line element) up to an overall factor: three
rotations, three translations, one dilatation and three special conformal
transformations. The first six of these are well-known for leaving $ds^{2}$
invariant -- they form the Euclidean group for 3-dimensional space (or,
equivalently the inhomogeneous orthogonal group, $ISO(3)$). The single
dilatation is a simple rescaling. In Cartesian coordinates it is just 
\[
x^{i}=e^{\lambda }y^{i} 
\]
where $\lambda $ is any constant. The special conformal transformations are
actually a second kind of translation, performed in inverse coordinates,
given by: 
\[
q^{i}=\frac{x^{i}+x^{2}b^{i}}{1+2b^{i}x_{i}+b^{2}x^{2}} 
\]
The inverse is given by: 
\[
x^{i}=\frac{q^{i}-q^{2}b^{i}}{1-2q^{i}b_{i}+q^{2}b^{2}} 
\]

Here we prove directly that this transformation has the required effect of
transforming the metric conformally, according to 
\begin{equation}
\eta _{ab}\rightarrow \left( 1-2b^{i}x_{i}+b^{2}x^{2}\right) ^{-2}\eta _{ab}
\label{Metric transformation}
\end{equation}
Under $any$ coordinate transformation the metric changes according to 
\[
g_{ij}(q)=\eta _{mn}\frac{\partial x^{m}}{\partial q^{i}}\frac{\partial x^{n}%
}{\partial q^{j}} 
\]
For the particular case of the special conformal coordinate transformation
we have 
\[
x^{i}=\frac{q^{i}-q^{2}b^{i}}{1-2q\cdot b+q^{2}b^{2}}=\frac{1}{\beta }\left(
q^{i}-q^{2}b^{i}\right) 
\]
and therefore 
\[
\frac{\partial x^{m}}{\partial q^{i}}=\frac{1}{\beta ^{2}}\left( \beta
\delta _{i}^{m}-2\beta
b^{m}q_{i}+2b^{2}q^{2}b^{m}q_{i}+2q^{m}b_{i}-2b^{2}q^{m}q_{i}-2q^{2}b^{m}b_{i}\right) 
\]
Substituting into the metric transformation we find 
\begin{eqnarray*}
g_{ij} &=&\eta _{mn}\frac{\partial x^{m}}{\partial q^{i}}\frac{\partial x^{n}%
}{\partial q^{j}} \\
\beta ^{4}g_{ij} &=&\beta ^{2}\eta _{ij}-2\beta
^{2}b_{i}q_{j}+2b^{2}q^{2}\beta b_{i}q_{j}+\beta 2q_{j}b_{i}-4\beta \left(
b\cdot q\right) b_{i}q_{j} \\
&&+4b^{2}q^{2}\left( b\cdot q\right) b_{i}q_{j}-4b^{2}q^{2}b_{i}q_{j}+4\beta
b^{2}q^{2}b_{i}q_{j}-4b^{2}q^{2}q^{2}b^{2}b_{i}q_{j} \\
&&+4b^{2}q^{2}\left( b\cdot q\right) b_{i}q_{j}+2\beta q_{i}b_{j}-2\beta
^{2}b_{j}q_{i}-4\beta \left( b\cdot q\right) q_{i}b_{j}+4q^{2}\beta
b^{2}q_{i}b_{j} \\
&&+2\beta b^{2}q^{2}b_{j}q_{i}+4b^{2}q^{2}\left( b\cdot q\right)
q_{i}b_{j}-4q^{2}b^{2}q^{2}b^{2}q_{i}b_{j}-4b^{2}q^{2}q_{i}b_{j} \\
&&+4q^{2}b^{2}\left( b\cdot q\right) q_{i}b_{j}-2b^{2}\beta
q_{i}q_{j}+4\beta ^{2}b^{2}q_{i}q_{j}-4b^{2}q^{2}\beta b^{2}q_{i}q_{j} \\
&&+4b^{2}\beta \left( b\cdot q\right) q_{i}q_{j}-4\beta
b^{2}q^{2}b^{2}q_{i}q_{j}+4b^{2}q^{2}b^{2}q^{2}b^{2}q_{i}q_{j} \\
&&-4b^{2}b^{2}q^{2}\left( b\cdot q\right) q_{i}q_{j}-2\beta
b^{2}q_{j}q_{i}+4\beta b^{2}\left( b\cdot q\right) q_{i}q_{j} \\
&&-4b^{2}q^{2}b^{2}\left( b\cdot q\right)
q_{i}q_{j}+4b^{2}b^{2}q^{2}q_{i}q_{j}-2q^{2}\beta b_{i}b_{j}+4q^{2}b_{i}b_{j}
\\
&&-4q^{2}\left( b\cdot q\right) b_{i}b_{j}-2\beta
q^{2}b_{j}b_{i}-4q^{2}\left( b\cdot q\right)
b_{i}b_{j}+4q^{2}q^{2}b^{2}b_{i}b_{j}
\end{eqnarray*}
The terms proportional to $q_{i}q_{j}$ cancel identically, while the
remaining terms coalesce into $\beta $ factors which cancel, leaving simply 
\[
g_{ij}=\beta ^{-2}\eta _{ij} 
\]
This is a direct proof that the given transformation is conformal.

It is possible to show that the ten transformations described here are the
only conformal transformations in 3 dimensions.

\pagebreak

\noindent {\large Appendix 3: What is the velocity after a special conformal
transformation?}

Suppose a particle follows the path $\mathbf{x}\left( t\right) $ with
velocity 
\[
\mathbf{v}=\frac{d\mathbf{x}\left( t\right) }{dt} 
\]
If we introduce new coordinates 
\[
\mathbf{y}=\frac{\mathbf{x}+x^{2}\mathbf{b}}{1+2\mathbf{x}\cdot \mathbf{b}%
+b^{2}x^{2}}=\beta ^{-1}\left( \mathbf{x}+x^{2}\mathbf{b}\right) 
\]
where $\beta =\left( 1+2\left( \mathbf{x}\cdot \mathbf{b}\right)
+b^{2}x^{2}\right) .$ Then 
\begin{eqnarray*}
\mathbf{y}\left( t\right) &=&\frac{\mathbf{x}\left( t\right) +x^{2}\left(
t\right) \mathbf{b}}{1+2\mathbf{x}\left( t\right) \cdot \mathbf{b}%
+b^{2}x\left( t\right) ^{2}} \\
\mathbf{x}\left( t\right) &=&\frac{\mathbf{y}-y^{2}\mathbf{b}}{1-2\mathbf{y}%
\cdot \mathbf{b}+b^{2}y^{2}}
\end{eqnarray*}
Then 
\begin{equation}
\frac{\partial y^{i}}{\partial x^{j}}=\beta ^{-1}\left( \delta
_{j}^{i}+2x_{j}b^{i}\right) -\beta ^{-2}\left( x^{i}+x^{2}b^{i}\right)
\left( 2b_{j}+2b^{2}x_{j}\right)  \label{Jacobian}
\end{equation}
This is just as complicated as it seems. The velocity in the new coordinates
is 
\begin{eqnarray}
\frac{dy^{i}}{dt} &=&\beta ^{-1}\left( \mathbf{v}+2\left( \mathbf{x}\cdot 
\mathbf{v}\right) \mathbf{b}\right) -\beta ^{-2}\left( \mathbf{x}+x^{2}%
\mathbf{b}\right) \left( 2\mathbf{v}\cdot \mathbf{b}+2b^{2}\left( \mathbf{x}%
\cdot \mathbf{v}\right) \right)  \nonumber \\
&=&v^{j}\left( \beta ^{-1}\left( \delta _{j}^{i}+2x_{j}b^{i}\right) -\beta
^{-2}\left( x^{i}+x^{2}b^{i}\right) \left( 2b_{j}+2b^{2}x_{j}\right) \right)
\label{Velocity transform}
\end{eqnarray}
The explicit form is probably the basis for Weinberg's claim \cite{Weinberg}%
, that under conformal transformations ``\ldots the statement that a free
particle moves at constant velocity [is] not an invariant statement....''
This is clearly the case -- if $\frac{dx^{i}}{dt}=v^{i}$ is constant, $\frac{%
dy^{i}}{dt}$ depends on position in a complicated way. However, we note that
using eq.(\ref{Jacobian}) we may rewrite eq.(\ref{Velocity transform}) in
the usual form for the transformation of a vector. 
\[
\frac{\partial y^{i}}{\partial x^{j}}=\frac{\partial y^{i}}{\partial x^{j}}%
v^{j} 
\]
This is the reason we must introduce a derivative operator covariant with
respect to special conformal transformations. The statement $%
v^{k}D_{k}v^{i}=0$ is then a manifestly conformally covariant expression of
constant velocity.

\pagebreak

\noindent {\large Appendix 4: The geometry of special conformal
transformations}

We have shown that 
\[
g_{ij}=\beta ^{-2}\eta _{ij} 
\]

But notice that, if we perform such a transformation, the connection and
curvature no longer vanish, but are instead given by 
\begin{eqnarray*}
\mathbf{e}^{a} &=&\beta ^{-1}\mathbf{d}x^{a} \\
\mathbf{de}^{a} &=&\mathbf{e}^{b}\mathbf{\omega }_{b}^{a} \\
\mathbf{R}_{b}^{a} &=&\mathbf{d\omega }_{b}^{a}-\mathbf{\omega }_{b}^{c}%
\mathbf{\omega }_{c}^{a}
\end{eqnarray*}
This system is simple to solve. From the second equation, we have 
\[
\mathbf{\omega }_{b}^{a}=-\left( \beta _{,b}\mathbf{e}^{a}-\eta ^{ac}\eta
_{bd}\beta _{,c}\mathbf{e}^{d}\right) 
\]
Then substituting into the curvature, 
\begin{eqnarray*}
\mathbf{R}_{b}^{a} &=&\mathbf{d\omega }_{b}^{a}-\mathbf{\omega }_{b}^{c}%
\mathbf{\omega }_{c}^{a} \\
&=&-\delta _{d}^{a}\beta _{,bc}\mathbf{e}^{c}\mathbf{e}^{d}+\eta ^{ae}\eta
_{bd}\beta _{,ec}\mathbf{e}^{c}\mathbf{e}^{d}+\delta _{d}^{a}\beta
_{,c}\beta _{,b}\mathbf{e}^{c}\mathbf{e}^{d}-\eta ^{ae}\eta _{bd}\beta
_{,c}\beta _{,e}\mathbf{e}^{c}\mathbf{e}^{d} \\
&&-\delta _{d}^{a}\beta _{,b}\beta _{,c}\mathbf{e}^{c}\mathbf{e}^{d}+\delta
_{d}^{a}\eta _{bc}\eta ^{fe}\beta _{,f}\beta _{,e}\mathbf{e}^{c}\mathbf{e}%
^{d}-\eta _{bc}\eta ^{af}\beta _{,d}\beta _{,f}\mathbf{e}^{c}\mathbf{e}^{d}
\\
R_{bcd}^{a} &=&\delta _{c}^{a}\beta _{,bd}-\delta _{d}^{a}\beta _{,bc}+\eta
^{ae}\eta _{bd}\beta _{,ec}-\eta ^{ae}\eta _{bc}\beta _{,ed} \\
&&+\left( \delta _{d}^{a}\eta _{bc}-\delta _{c}^{a}\eta _{bd}\right) \eta
^{fe}\beta _{,f}\beta _{,e}
\end{eqnarray*}
which is pure Ricci. We knew this in advance because the Weyl curvature
tensor vanishes for conformally flat metrics. The Ricci tensor and Ricci
scalar are 
\begin{eqnarray*}
R_{bd} &=&\left( n-2\right) \beta _{,bd}+\eta _{bd}\eta ^{ce}\beta
_{,ec}-\left( n-1\right) \eta _{bd}\eta ^{fe}\beta _{,f}\beta _{,e} \\
R &=&2\left( n-1\right) \eta ^{bd}\beta _{,bd}-n\left( n-1\right) \eta
^{fe}\beta _{,f}\beta _{,e}
\end{eqnarray*}
where 
\begin{eqnarray*}
\partial _{a}\beta &=&\partial _{a}\left( 1-2x\cdot b+x^{2}b^{2}\right) \\
&=&-2b_{a}+2b^{2}x_{a} \\
\partial _{ab}\beta &=&\partial _{b}\left( -2b_{a}+2b^{2}x_{a}\right) \\
&=&2b^{2}\eta _{ab}
\end{eqnarray*}
so finally, 
\begin{eqnarray*}
R_{bd} &=&4\left( n-1\right) b^{2}\left( 1-\beta \right) \eta _{bd} \\
R &=&4n\left( n-1\right) b^{2}\left( 1-\beta \right)
\end{eqnarray*}
The full curvature therefore is determined fully by the Ricci scalar: 
\begin{eqnarray*}
R_{bcd}^{a} &=&\left( \delta _{c}^{a}\eta _{bd}-\delta _{d}^{a}\eta
_{bc}\right) 4b^{2}\left( 1-\beta \right) \\
&=&\frac{R}{n\left( n-1\right) }\left( \delta _{c}^{a}\eta _{bd}-\delta
_{d}^{a}\eta _{bc}\right)
\end{eqnarray*}
where 
\[
R=4n\left( n-1\right) b^{2}\left( 2x\cdot b-x^{2}b^{2}\right) 
\]

\pagebreak

\noindent {\large Appendix 5: The }$ISO(3)${\large \ and conformal }$SO(4,1)$%
{\large \ Lie algebras}

For our gauging, we require the form of the Lie algebras $iso\left( 3\right) 
$ and $so\left( 4,1\right) .$ We can find both from the general form of any
pseudo-orthogonal Lie algebra. Let $\eta _{AB}=diag(1,\ldots ,1,-1,\ldots
,-1)$ with $p$ $+1s$ and $q$ $-1s.$ Then the Lie algebra $o(p,q)$ is, up to
normalization, 
\[
\left[ M_{AB},M_{CD}\right] =\eta _{BC}M_{AD}-\eta _{BD}M_{AC}-\eta
_{AC}M_{BD}+\eta _{AD}M_{BC} 
\]

The Lie algebra $iso(3)$ may be found as a contraction of $o(4)$ or $o(3,1).$
Let $\eta _{AB}=diag(1,1,1,1)$, let $\eta _{ij}=diag\left( 1,1,1\right) ,$
and replace $M_{i4}$ and $M_{4i}$ by 
\[
\lambda P_{i}=M_{i4}=-M_{4i} 
\]
Separating the $i=1,2,3$ parts from the $i=4$ parts, 
\begin{eqnarray*}
\left[ M_{ij},M_{kl}\right] &=&\eta _{jk}M_{il}-\eta _{jl}M_{ik}-\eta
_{ik}M_{jl}+\eta _{il}M_{jk} \\
\left[ M_{ij},\lambda P_{k}\right] &=&\eta _{jk}\lambda P_{i}-\eta
_{ik}\lambda P_{j} \\
\left[ \lambda P_{i},\lambda P_{k}\right] &=&-\eta _{44}M_{ik}
\end{eqnarray*}
so in the limit as $\lambda \rightarrow \infty ,$%
\begin{eqnarray*}
\left[ M_{ij},M_{kl}\right] &=&\eta _{jk}M_{il}-\eta _{jl}M_{ik}-\eta
_{ik}M_{jl}+\eta _{il}M_{jk} \\
\left[ M_{ij},P_{k}\right] &=&\eta _{jk}P_{i}-\eta _{ik}P_{j} \\
\left[ P_{i},P_{j}\right] &=&-\frac{1}{\lambda ^{2}}M_{ik}\rightarrow 0
\end{eqnarray*}
While these relations hold in any dimension, in 3-dim we can simplify the
algebra using the Levi-Civita tensor to write 
\begin{eqnarray*}
J_{i} &=&-\frac{1}{2}\varepsilon _{i}^{\quad jk}M_{jk} \\
M_{ij} &=&-\varepsilon _{ij}^{\quad k}J_{k}
\end{eqnarray*}
Then 
\begin{eqnarray*}
\left[ J_{m},J_{n}\right] &=&\varepsilon _{mn}^{\quad \quad i}J_{i} \\
\left[ J_{i},P_{j}\right] &=&\varepsilon _{ij}^{\quad n}P_{n} \\
\left[ P_{i},P_{j}\right] &=&0
\end{eqnarray*}

The conformal Lie algebra is just $o(4,1),$ given by 
\[
\left[ M_{AB},M_{CD}\right] =\eta _{BC}M_{AD}-\eta _{BD}M_{AC}-\eta
_{AC}M_{BD}+\eta _{AD}M_{BC} 
\]
with $A,B,\ldots =1,2,\ldots ,5.$ To see this algebra in terms of the usual
definite-weight generators we write the metric in the form 
\[
\eta _{AB}=\left( 
\begin{array}{ccccc}
1 &  &  &  &  \\ 
& 1 &  &  &  \\ 
&  & 1 &  &  \\ 
&  &  & 0 & 1 \\ 
&  &  & 1 & 0
\end{array}
\right) =\eta ^{AB} 
\]
The Lie algebra of generators $M_{\quad b}^{a},P_{a},K^{a}$ and $D$ is then
found by setting 
\begin{eqnarray*}
P_{a} &=&M_{\quad a}^{4} \\
K^{a} &=&\eta ^{ab}M_{\quad b}^{5} \\
D &=&M_{\quad 4}^{4}=-M_{\quad 5}^{5}
\end{eqnarray*}
With $a,b,\ldots =1,2,3$ we find 
\begin{eqnarray}
\left[ M_{\quad b}^{a},M_{\quad d}^{c}\right] &=&\delta _{b}^{c}M_{\quad
d}^{a}-\eta ^{ca}\eta _{be}M_{\quad d}^{e}-\eta _{bd}M^{ac}+\delta
_{d}^{a}M_{b}^{\quad c}  \nonumber \\
\left[ M_{\quad b}^{a},P_{c}\right] &=&\eta _{bc}\eta ^{ae}P_{e}-\delta
_{c}^{a}P_{b}  \nonumber \\
\left[ M_{\quad b}^{a},K^{c}\right] &=&\delta _{b}^{c}K^{a}-\eta ^{ca}\eta
_{be}K^{e}  \nonumber \\
\left[ P_{b},K_{d}\right] &=&-\eta _{be}M_{\quad d}^{e}-\eta _{bd}D 
\nonumber \\
\left[ D,P_{a}\right] &=&-P_{a}  \nonumber \\
\left[ D,K^{a}\right] &=&K^{a}
\end{eqnarray}
This is the usual form of the conformal algebra. The matrices $M_{\quad
b}^{a}$ generate $SO(3),$ the three generators $P_{a}$ lead to translations, 
$K^{a}$ give translations of the point at infinity (special conformal
transformations), and $D$ generates dilatations.

\end{document}